\documentclass[aps,prd,twocolumn,floatfix,a4paper,showpacs,showkeys,nofootinbib]{revtex4-1}
\usepackage{amsmath,amssymb,bm,graphicx,geometry,mathtools,array,multirow,slashed,etoolbox,upgreek,mathrsfs,hyperref}
\usepackage{soul}
\allowdisplaybreaks
\makeatletter
\newcommand{\vast}{\bBigg@{4}}
\newcommand{\Vast}{\bBigg@{5}}
\makeatother
\begin{document}
\title{Constraints imposed by the partial wave amplitudes on the decays of \texorpdfstring{$J=1,2$}{J=1,2} mesons}
\author{Vanamali Shastry}
\email{vanamalishastry@gmail.com}
\affiliation{Institute of Physics, Jan Kochanowski University, ul.  Uniwersytecka  7,  P-25-406  Kielce,  Poland}
\author{Enrico Trotti}
\email{trottienrico@gmail.com}
\affiliation{Institute of Physics, Jan Kochanowski University, ul.  Uniwersytecka  7,  P-25-406  Kielce,  Poland}
\author{Francesco Giacosa}
\email{fgiacosa@ujk.edu.pl}
\affiliation{Institute of Physics, Jan Kochanowski University, ul.  Uniwersytecka  7,  P-25-406  Kielce,  Poland}
\affiliation{Institute for Theoretical Physics, Johann Wolfgang Goethe - University, Max von Laue--Str. 1 D-60438 Frankfurt, Germany}

%\maketitle
\begin{abstract}
    We study the two-body decays of mesons using the covariant helicity formalism. In particular, we show how the partial wave analysis of decays constrains the interacting terms entering the Lagrangian describing the decays of mesons with $J=1$ and $J=2$.  We use available information on partial wave analysis to study specific mesonic decays and to make predictions for not yet measured quantities as well as to investigate the isoscalar mixing angle in the axial-vector, pseudovector and pseudotensor sectors. In particular, in the axial-vector sector our result  agrees with the LHCb one, and in the pseudotensor sector we confirm  a quite large (and negative) angle in the nonstrange-strange basis, which is compatible with a large contribute of the axial anomaly. 
\end{abstract}
\maketitle
\section{Introduction}\label{intro}
The study of mesons and their decays can provide a wealth of information
regarding the interactions between the various states as well as the
internal dynamics of the states involved, and ultimately, the strong interactions.

On the experimental front, a lot of effort has been dedicated to the study of mesonic
decays (e.g. Refs.
\cite{Zyla:2020zbs,bes,compass,COMPASS:2009xrl,lhcb,gluex1,gluex2,panda,Amelino-Camelia:2010cem}), as these
decays are a way to generate and observe new states as well as portals to
possible new physics. On the theoretical front too, a wealth of knowledge has
been gained using various field theoretic models, quarks models, and
effective field theories
\cite{Godfrey:1985xj,Isgur:1984bm,Klempt:2007cp,Pelaez:2015qba,lutz}.

A vast majority of the phenomenological models formulated till date estimate
the coupling constants by analyzing the mass of the mesons, their widths, and the
branching fractions of their decays. One crucial set of data points available
in the PDG, the ratio of the partial wave amplitudes (PWAs, see for instance Ref. \cite{Peters:2004qw}), 
is usually not taken into account. We demonstrate in the present work how this particular data can be
used to build more robust field theoretic phenomenological models and to put a tighter
constraint on their parameters.

In order to set the frame of our work, let us consider a decay of the type
\[
A\rightarrow BC\text{ ,}%
\]
where $A,$ $B,$ $C$ are certain mesonic fields with definite total spin
$J_{A},$ $J_{B},$ and $J_{C}.$ The corresponding interaction Lagrangian that
describes this decay process should fulfill the basic constraint such as
Lorentz as well as, for QCD processes, parity and charge-conjugation
invariance. It can be expressed as%
\begin{equation}
\mathcal{L}_{ABC}=\mathcal{L}_{ABC}^{\text{c}}+\mathcal{L}_{ABC}^{\text{d}%
}+\ldots\label{lag}%
\end{equation}
where $\mathcal{L}_{ABC}^{\text{c}}$ contains the lowest possible number of
derivatives, while $\mathcal{L}_{ABC}^{\text{d}}$ is the next term with two additional derivatives, etc. 

Various approaches, based on the realization of flavor symmetry or, more
generally, on the linear realization of chiral symmetry consider the first
term as dominant, e.g. Refs.
\cite{Koenigstein:2015asa,Divotgey:2013jba,dick,korudaz,carter,fariborz},
while approaches based on the non-linear realization of chiral symmetry
typically contain terms with higher derivatives (see for instance, Refs.
\cite{chpt,chptvm,Bernard:2006gx,Jenkins:1995vb,Booth:1996hk,cirigliano,Terschlusen:2016kje}). 

An important aspect of the decays $A\rightarrow B+C$ is that different waves
for the final product are possible. Denoting with $\ell$ the relative orbital
angular momentum between $B$ and $C,$ the possible values of $\ell$ range between
$|J_{A}-(J_{B}+J_{C})|$ up to $J_{A}+(J_{B}+J_{C})$. For
instance, in the decay $a_{1}(1260)\rightarrow\rho\pi$ the waves $\ell=0$ and
$\ell=2$ are allowed, while in the $\pi_{2}(1670)\rightarrow f_{2}(1270)\pi$ decay one may
have $\ell=0,2,4.$ The ratio between two allowed $\ell$-values can be determined by an appropriate PWA analysis \cite{Zyla:2020zbs}.

A natural question regards the connection of the interaction terms in Eq.
(\ref{lag}) to the ratio of partial waves. In general, each interaction
Lagrangian gives a nonzero contribution to each partial wave. For instance,
one may ask for a certain decay if the term with the lowest number of
derivatives is sufficient do describe data or not. Conversely, the possibility
that the derivative interaction term dominates can be also addressed. 

In this work, we study, in a systematic framework, the PWA of the decays of the axial-vector, pseudovector, and pseudotensor mesons by using model Lagrangian(s) of the type of Eq. (\ref{lag}). Our aim is to understand the role played by the various
interactions in the decays of these mesons. Within this respect, the information gained by PWA turns out to be very useful. We thus can -on the one hand- reproduce previous results on the subject (in particular Ref. \cite{Jeong:2018exh}, in which also a model Lagrangian was used), and on the other hand, extend the procedure to the whole class of unstable high-spin mesons mentioned above. 
Moreover, we shall analyze (to our knowledge for the first time using PWA) the mixing in the isoscalar sector of the investigated mesonic nonets. Namely, the question about the role of the anomaly, besides the well-known case of the pseudoscalar sector, is on its own an interesting aspect of nonperturbative QCD \cite{tHooft:1986ooh,Christos:1984tu,Giacosa:2017pos}.\par

Our results about the PWA shall be compared to those of other approaches, such as the PWA analysis of the $^3P_0$ model and the lattice calculations. We find that our results agree with those of the $^3P_0$ model in the $J=1$ sector \cite{Barnes:1996ff}. On the lattice front, the $b_1(1235)\to\omega\pi$ decay was studied by the Hadron Spectrum collaboration recently \cite{Woss:2019hse}. Their inference that the $b_1(1235)$ couples strongly to the $S-$wave $\omega\pi$ compared to the $D-$wave is in line with experimental results \cite{Zyla:2020zbs}.\par

This paper is organized into four sections. In Sec. \ref{fPWA}, we discuss the
formalism used in deriving the PWAs and the construction of the polarization
tensors. In Sec. \ref{derPWA} we derive the partial wave amplitudes for the
different decays discussed in the paper, and analyze their behavior. In Sec.
\ref{RnDPWA}, we discuss the results of the work and their consequences.
Finally, we summarize the entire work in Sec. \ref{SnCPWA}.

\section{Partial Wave Amplitudes}\label{fPWA}
Much research has been conducted on the partial wave decomposition of the decay processes. One of the earliest works in this direction was the tensor formalism by Zemach \cite{Zemach:1968zz,Zemach:1963bc}. In this formalism, the decay amplitude is written in terms of the non-covariant 3-dimensional spin tensors defined in the rest frame of each decaying particle. This results in a {\it frame dependent} decay width which leads to hurdles in interpreting the square of the amplitude as the decay probability. 

An alternative approach to analyzing the partial waves is the helicity formalism. Initiated by Jacob and Wick \cite{Jacob:1959at}, the helicity formalism has been used extensively to study the decay processes. In this formalism, the angular dependence of the decay process is captured in the Wigner D-matrices $D^J_{mm^\prime}$. The remaining part of the decay amplitude forms the helicity coupling amplitude. In a typical scenario, where experimental data has to be analysed, the helicity amplitudes are constructed empirically using the Breit-Wigner functions and the centrifugal functions - which are nothing but the moduli of the Zemach tensors. This approach makes the formalism non-covariant, making it unsuitable for practical applications as the decay amplitude must be a Lorentz invariant.\par
Chung proposed a covariant form of the helicity formalism in which the helicity coupling amplitude is constructed from the polarization tensors and hence is a function of the ratio $E/m$ ($E$ and $m$ are the energy and mass of the particles involved in the decay process, as measured in the rest frame of the parent) making it a Lorentz scalar \cite{Chung:1993da,Chung:1997jn}. In the present work, we make use of model Lagrangians to write down the amplitude of the decays. We then derive the  helicity coupling amplitudes from the decay amplitudes which we find to be functions of the energy (or $3-$momentum) of the daughter mesons and rest masses of the mesons involved, as measured in the rest frame of the parent.\par
In the following subsection, we discuss briefly the covariant helicity formalism.\par

\subsection{The covariant helicity formalism}
Consider the two-body decay process, $A\to BC $. Let the total angular momentum states of the particles $A\,,\, B\,,\text{ and }C$ be $|J,M_J\rangle\,,\, |s,\lambda\rangle\,,\text{ and }|\sigma\,,\nu\rangle$ respectively. Also, let the sum of the total spin quantum numbers of the daughter states be given by $S$, {\it i.e,} 
\begin{equation}
    |S,m_s\rangle = |s,\lambda\rangle\oplus|\sigma\,,\nu\rangle,
\end{equation} 
where $\oplus$ implies that the $ |S,m_s\rangle$ state is constructed from $|s,\lambda\rangle \text{ and }|\sigma,\nu\rangle$ by following the rules of addition of the angular momenta. The spin of the parent can then be constructed by adding the total spin of the daughters with the relative orbital angular momentum ($\ell$) carried by them, {\it i.e,}
\begin{align}
    |J,M_J\rangle &= |\ell,m_\ell\rangle \oplus |S,m_s\rangle \text{ .}
\end{align}
Thus, unlike a two-body scattering process where an infinity of angular momentum channels are available, the number of angular momentum channels available for a two-body decay is limited by the spins of the parent and daughter states. The value of $\ell$ must satisfy the condition that $J\in[|\ell-S|,~\ell+S]$. Also, since we are interested only in the strong decays, an additional constraint of parity conservation has to be imposed. This determines if $\ell$ has to be even or odd (for a given value of $S$), further reducing the available number of angular momentum channels.

The amplitude for a two-body decay can be written as
\begin{equation}
\mathcal{M}^J(\theta,\phi\,;M_J) \propto D^{J\ast}_{M\delta}(\phi,\theta,0)F^J_{\lambda\nu},\label{amp1hel}
\end{equation}
where $D^{J\ast}_{M\delta}(\phi,\theta,0)$ is the complex conjugate of the Wigner $D-$matrix, $ F^J_{\lambda\nu}$ is the helicity amplitude, and $\delta=\lambda-\nu$. Eq. (\ref{amp1hel}) is a general result, and any model dependence will appear in the exact form of the helicity amplitudes. As a special case, if the frame of reference is such that the decay products are aligned along the $\pm z-$axis, the decay amplitude becomes proportional to only the helicity amplitude:
\begin{equation}
\mathcal{M}^J_{\lambda\delta}(0,0\,;M_J) \propto F^J_{\lambda\nu}.\label{amp2hel}
\end{equation}
When the decay products are massive, the helicity amplitudes can be expanded in terms of the $\ell S$ coupling amplitudes ($G^J_{\ell S}$) through the relation
\begin{equation}
F^J_{\lambda\nu} = \sum_{\ell S} \sqrt{\frac{2\ell+1}{2J+1}} \langle\ell 0 S\delta|J\delta\rangle\langle s\lambda\sigma-\nu|S\delta\rangle G^J_{\ell S},\label{amp1LS}
\end{equation}
where $\langle\cdots|\cdots\rangle$ represent the Clebsch-Gordan coefficients. As explained above, the allowed values of $\ell$ are determined by the spin and parity of the parent and the decay products. 

Some comments regarding the validity of the above relation are in order. Firstly, the $\ell S$ coupling amplitudes can be chosen in two ways: (i) empirically, by using the rule $G_{\ell S}^J \propto |\vec{k}|^\ell$, where $|\vec{k}|$ is the magnitude of the break-up momentum; (ii) from the polarization vectors. In the former case, the helicity amplitudes become non-covariant due to the frame dependence introduced by the choice of $G^J_{\ell S}$. The helicity amplitudes can be made Lorentz scalar using the latter method, if the polarization vectors are boosted to the appropriate frame \cite{Chung:1997jn,Filippini:1995yc}.
The ratio of $G_{\ell S}^J$ gives us the ratio of the partial wave amplitudes.\par
Alternatively, one can expand the decay amplitude in terms of the spherical harmonics as
\begin{align}
    i\mathcal{M}(\theta,\phi;M_J)\! &=\! i \sum_\ell\!\! \sum_{m_\ell=-\ell}^\ell\!\! G_\ell \langle \ell m_\ell S m_s | JM_J \rangle Y_{\ell m_\ell}(\theta,\phi).
\end{align}
The PWAs so derived will be proportional to the PWAs derived using the covariant helicity formalism {\it i.e,} 
\begin{align}
    G_\ell &= \sqrt{\frac{\alpha}{(2J+1)}}G^J_{\ell S} ,   
\end{align}
where $\alpha$ is a numerical factor dependent on the normalization of the spherical harmonics. For the normalization $\int d\Omega |Y_{\ell m_\ell}|^2 = 1$, $\alpha=4\pi$. The advantage of using the covariant helicity formalism is that, by choosing the helicity amplitudes suitably, we obtain
\begin{align}
\sum_{\ell S} |G^J_{\ell S}|^2 = \sum_{\rm spins}|\mathcal{M}|^2.
\end{align}

\subsection{Polarization states}
The present study is concerned with the decay of mesons with $J\ge1$ in to states with one of them having $J\ge1$. We detail the construction of the polarization vectors (PV) and polarization tensors (PT) in this subsection.\par
The PVs of a spin$-1$ state in its rest frame are given by
\begin{eqnarray}
\epsilon^\mu(\vec{0},+1)&=-\frac{1}{\sqrt{2}}\begin{pmatrix}0, 1, i, 0\end{pmatrix}\,,\nonumber\\
\epsilon^\mu(\vec{0},-1)&=\frac{1}{\sqrt{2}}\begin{pmatrix}0,1,-i,0\end{pmatrix}\,,\nonumber\\
\epsilon^\mu(\vec{0},0)&=\begin{pmatrix}0,0,0,1\end{pmatrix}.
\end{eqnarray}
These PVs satisfy the following orthonormality conditions:
\begin{align}
    k_\mu \epsilon^\mu(\vec{k},m)&=0\\
    \epsilon^\ast_\mu(\vec{k},m)\epsilon^\mu(\vec{k},m^\prime)&=-\delta_{mm^\prime}.
\end{align}
Further, the projection operator is given by the identity
\begin{align}
    \tilde{g}_{\mu\nu}= \sum_m \epsilon_\mu(\vec{k},m) \epsilon^\ast_\nu(\vec{k},m)&= -g_{\mu\nu} + \frac{k_\mu k_\nu}{M_0^2},\label{projop}
\end{align}
where $k_\mu$ and $M_0$ are the 4-momentum and mass of the corresponding state respectively. The PTs for higher spin states can be constructed from the PVs using a standard algorithm. The PTs for a spin-$J$ state can be constructed using the master formula
\begin{widetext}
\begin{align}
    \epsilon^{\mu_1\mu_2\ldots\mu_J}(\vec{0},m)\! &=\!\! \sum_{m_1m_2\ldots}\!\! \langle 1m_1 1m_2| 2n_1 \rangle \langle 2 n_1 1m_3 | 3n_2 \rangle \ldots \langle J-1 n_{J-2} 1m_J | J m \rangle\nonumber\\ &\qquad\otimes\epsilon^{\mu_1}(\vec{0},m_1)\epsilon^{\mu_2}(\vec{0},m_2)\ldots \epsilon^{\mu_J}(\vec{0},m_J).
\end{align}
\end{widetext}
The states constructed using this algorithm satisfy the following orthonormality relations:
\begin{align}
    k_{\mu_i}\epsilon^{\mu_1\mu_2\ldots\mu_J}(m)&=0\\
    \epsilon^\ast_{\mu_1\mu_2\ldots\mu_J}(m)\epsilon^{\mu_1\mu_2\ldots\mu_J}(m^\prime) &= (-1)^J\delta_{mm^\prime}
\end{align}
and transform under rotations as
\begin{align}
   \!\!\!\!\epsilon^{\mu_1\mu_2\ldots\mu_J}(m)\to \sum_m{^\prime}\epsilon^{\mu_1\mu_2\ldots\mu_J}(m^\prime) D^J_{m^\prime m}(\phi,\theta,\psi).
\end{align}
 We list below the explicit expressions for spin$-2$ states:
\begin{widetext}
\begin{align}
    \epsilon^{\mu\nu}(\vec{k},+2) &= \epsilon^\mu(\vec{k},+1)\epsilon^\nu(\vec{k},+1)\\
    \epsilon^{\mu\nu}(\vec{k},+1) &= \frac{1}{\sqrt{2}}\big[\epsilon^\mu(\vec{k},+1)\epsilon^\nu(\vec{k},0) + \epsilon^\mu(\vec{k},0)\epsilon^\nu(\vec{k},+1) \big]\\
    \epsilon^{\mu\nu}(\vec{k},0) ~&= \frac{1}{\sqrt{6}}\big[ \epsilon^\mu(\vec{k},+1)\epsilon^\nu(\vec{k},-1) + \epsilon^\mu(\vec{k},-1)\epsilon^\nu(\vec{k},+1)\big] + \sqrt{\frac{2}{3}} \epsilon^\mu(\vec{k},0)\epsilon^\nu(\vec{k},0).
\end{align}
\end{widetext}
The PTs for states with $m=-1,-2$ can be obtained similarly by using the PVs with $m=-1$. The above definitions are valid for any particle moving with any 3-momentum $\vec{k}$ provided that the corresponding PVs are boosted appropriately before arriving at the PTs. Alternatively, one can construct the PTs using the PVs defined in the rest frame of the meson, and then boost the resultant PTs to the required frame.\par

\section{Deriving the \texorpdfstring{PWA\lowercase{s}}{PWAs}}\label{derPWA}
In this section, we derive the PWAs for three decay processes {\it viz.,} $a_1(1260)\to \rho\pi$, $\pi_2(1670)\to f_2(1270)\pi$, and $\pi_2(1670)\to\rho\pi$. In principle, the following discussions can be extended to the decays of all the members of the corresponding nonets. The general results derived for the decay of the $a_1(1260)$ can also be extended to the decays of the $b_1(1235)$ meson.
\subsection{The \texorpdfstring{$a_1(1260)\to\rho\pi$}{} decay}\label{demoa1}
The decay of the $a_1(1260)$ to $\rho\pi$ can be represented by the Lagrangian
\begin{align}
    \mathcal{L}&= ig_c^A \langle a_{1,\mu}\rho^\mu \pi\rangle + ig_d^A\langle\mathfrak{a}_{1,\mu\nu}\uprho^{\mu\nu}\pi\rangle,\label{lagJ1}
\end{align}
where $g_c^A$ and $g_d^A$ are the coupling constants, and $\mathfrak{a}_{1,\mu\nu}=\partial_\mu a_{1,\nu}-\partial_\nu a_{1,\mu}$ ,$\uprho^{\mu\nu}=\partial^\mu \rho^{\nu}-\partial^\nu \rho^{\mu}$ and $\langle~\rangle$ represents trace over the isospin. The Lagrangian consists of two types of interactions\footnote{Here, and in the following, we use the term ``contact interactions" or ``local interactions" to refer to operators without derivatives. Conversely, we  call ``derivative interactions" or ``nonlocal interactions" for the other terms.}: local (contact) interactions and nonlocal (derivative) interactions. As we discuss in a while, the local interactions are sufficient to reproduce the $D/S-$ratio of the $a_1(1260)\to\rho\pi$ decay. We write down the full amplitude (including both interactions) as
\begin{align}
i\mathcal{M} &=  g_c^A~ \epsilon_\mu(0,M_J)\epsilon^{\mu\ast}(\vec{k},\lambda)\nonumber\\
& +  2g_d^A \big[ k_0\cdot k_1 ~ \epsilon^\mu(\vec{0},M_J)\epsilon^\ast_\mu(\vec{k_1},\lambda)\nonumber\\
& - k_0^\nu~ k_{1,\mu}~\epsilon^\mu(\vec{0},M_J)\epsilon^\ast_\nu(\vec{k_1},\lambda)\big]\nonumber\label{ampJ1}\\
    &= -\Bigg\{ \begin{matrix*}[l]g_c^A + 2g_d^A~\!M_{a_1}~\!E_\rho& M_J=\lambda=\pm 1\\ \gamma(g_c^A + 2g_d^A~\! M_{a_1}~\! E_\rho\\ - 2g_d^A~\!M_{a_1}~\! \beta~\! k)& M_J=\lambda=0  \end{matrix*},
\end{align}
where $k_0^\mu=(M_{a_1},\vec{0})$ is the 4-momentum of the decaying meson, $k_1^\mu=(E_\rho,0,0,k)$ is the 4-momentum of the vector decay product, $M_{a_1}$ is the mass of the decaying meson, $M_\rho$ and $E_\rho$ are the mass and energy of the vector decay product respectively, and $k$ is the magnitude of the 3-momentum carried by the vector decay product. Notice that the last term in the Eq. (\ref{ampJ1}) contributes only when $M_J<|J|$. This statement is true for all the deays we have studied in this paper. The momentum dependence of the amplitude comes from the interaction terms as well as the polarization vectors. Thus, a simple Lagrangian with only contact interactions can also give rise to higher angular momentum partial waves in the amplitude, evern though these higher partial waves will be suppressed. Conversely, derivative interaction may also lead to lowest-order partial wave contributions.\par
\begin{figure}[ht]
\flushleft
\includegraphics[scale=0.85]{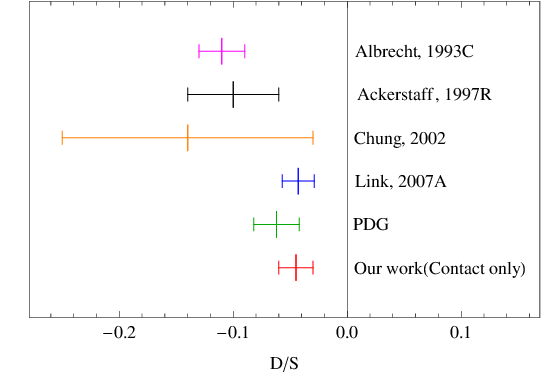}
\caption{$D/S-$ratio for the $a_1(1260)\to\rho\pi$ decay from various experiments and the uncertainties as listed in the PDG. For the sources of the values, see text.}\label{compDSa1}
\end{figure}

We now proceed with the analysis of the $a_1(1260)\to\rho\pi$ decay. The permitted values for the angular momentum quantum number are $\ell=0$ and $2$. Hence, from Eq. (\ref{amp1hel}),
\begin{align}
F^1_{10}&=  \frac{1}{\sqrt{3}} G_0 + \frac{1}{\sqrt{6}} G_2\\
F^1_{00}&=\frac{1}{\sqrt{3}} G_0 - \sqrt{\frac{2}{3}} G_2,
\end{align}
where $G_0 = G^1_{01}$ and $G_2=G^1_{21}$. We note that, if the helicity amplitudes $F_{10}^1 = F_{00}^1$, then the decay is entirely due to the $S-$wave, and if $F_{00}^1=2F_{10}^1$, the the decay is entirely due to $D-$wave.
From the amplitude (Eq. (\ref{ampJ1})), we see that
\begin{align}
F^1_{10}&=(g_c^A+2g_d^A M_{a_1} E_\rho)\\
F^1_{00}&=\gamma(g_c^A+2g_d^A~\! M_{a_1}~\! E_\rho - 2g_d^A M_{a_1}\beta k)
\end{align}
(up to a common multiplier). Now, we can invert the above relations to get the PWAs as\footnote{Here, and everywhere else, the partial wave amplitudes are derived up to an overall phase since the ratios of the PWAs do not depend on them.}
\begin{align}
    G_2 &= \sqrt{\frac{2}{3}}\left[ g_c^A\left(\frac{M_\rho - E_\rho}{M_\rho}\right) + 2g_d^A M_{a_1}( E_\rho - M_\rho)\right]\label{avG2}\\
    G_0 &= \frac{1}{\sqrt{3}}\left[ g_c^A\left(\frac{2M_\rho+E_\rho}{M_\rho}\right)+ 2g_d^A M_{a_1}(2E_\rho + M_\rho)\right]\label{avG0}.
\end{align}
If we ignore the derivative interactions ($g_d^A=0$), the ratio of the PWAs for the decay of $a_1(1260)\to\rho\pi$ is
\begin{align}
    \frac{G_2}{G_0} &= \sqrt{2}\left(\frac{M_\rho-E_\rho}{2M_\rho+E_\rho}\right).
\end{align}
Since in any frame of reference other than the rest frame of the $\rho-$meson $E_\rho>M_\rho$, $G_2$ is negative, and hence $|G_2|<|G_0|$. Substituting the values of the masses of the mesons involved and the magnitude of the 3-momentum carried by the decay products, we get $G_2/G_0=-0.045$, which is in good agreement with the value reported by the FOCUS collaboration \cite{FOCUS:2007ern} and within the error margin of the PDG value of $-0.062\pm 0.02$ \cite{Zyla:2020zbs}. In Fig. \ref{compDSa1}, we have shown the $D/S-$ratio for the $a_1(1260)\to\rho\pi$ decay obtained by various experiments along with the uncertainties compared to our work and the PDG average. Even when the derivative interactions are absent, our value is within the error estimate of the value obtained by the E852 collaboration (``Chung, 2002" \cite{Chung:2002pu}), and is in good agreement with that from the FOCUS experiment (``Link, 2007A" \cite{FOCUS:2007ern}). The values extracted by the OPAL collaboration (``Ackerstaff, 1997R" \cite{OPAL:1997was}) and the ARGUS (``Albrecht, 1993C" \cite{ARGUS:1992olh}) collaborations are significantly larger than our value. The experimental values and the corresponding uncertainties differ from each other significantly, as can be seen in Fig. \ref{compDSa1}. Hence, through out this study, we have used the PDG averages to estimate the parameters wherever needed.\par
The $a_1(1260)$ is a broad state with a width of $250 - 600$ MeV \cite{Zyla:2020zbs}. Since it is close to the $\rho\pi$ threshold, the width of the unstable $\rho-$meson can significantly influence the width of the $a_1(1260)\to\rho\pi$ decay. This can be estimated by performing a spectral integration of the decay width. However, we find that the decay width does not change the qualitative picture (see Appendix \ref{specintappe} for details).\par

Finally, the decay width is given by
\begin{align}
     \Gamma_{a_1\to\rho\pi} &= f_{a_1\rho\pi}\frac{k}{24\pi M_{a_1}^2}\Bigg[ (g_c^A)^2 \left(\frac{k^2}{M_\rho^2}+3\right)\nonumber\\
     &+ 12 g_c^A g_d^A E_\rho M_{a_1}\nonumber\\
     &+ 4(g_d^A)^2 M_{a_1}^2 M_\rho^2 \left(\frac{2k^2}{M_\rho^2}+3\right) \Bigg],\label{decwid1plus}
\end{align}
where $f_{a_1\rho\pi}$ is the isospin symmetry factor. The decay widths for the other members of the nonet can be obtained by using the corresponding values for the masses, energy, and isospin symmetry factor. The first term in the decay width arises purely from the contact interactions. The third term arises from the derivative interactions and adds to the contributions from the contact interactions. The second term is the interference between the contact and derivative interactions. The sign of $g_d^A$ indicates that the contact and derivative interactions interfere destructively.\par

The ratios of the PWAs for the decays of the pseudovector mesons can be calculated via the expressions given in Eq. (\ref{avG2}) and Eq. (\ref{avG0}) by using the appropriate masses and energies. The $b_1(1235)\to\omega\pi$ decay is comparable to the $a_1(1260)\to\rho\pi$ decay, in that, the masses of the mesons involved and the 3-momenta carried by the decay products are nearly equal. Thus, one would expect the ratios of the PWAs to be nearly the same for both the decays. The Lagrangian with only contact interactions when extended to the $b_1(1235)\to\omega\pi$ decay, in fact, gives the value of the ratio as $-0.043$. However, the experimentally observed value is much different at $0.277\pm 0.027$ \cite{Zyla:2020zbs}. 

This discrepancy can be addressed by including nonlocal interactions in the Lagrangian. The observed value of the magnitude of the $D/S-$ratio for the $b_1(1235)\to\omega\pi$ decay can be explained if the coupling constants have the ratio $g_d^B/g_c^B = -0.659$ GeV$^{-2}$, as given the Table \ref{1pluspar} (see also the discussions in Sec \ref{RnDPWA}). We observe that this ratio is very close to $1/M_{b_1}^2$ in magnitude. Such a relation between the ratio of the coupling constants and the mass of the decaying state occurs in all the decays we have studied in this paper.\par
\subsection{The \texorpdfstring{$\pi_2(1670)\to f_2(1270)\pi$}{} decay}\label{demopi2}
 
We introduce the following Lagrangian to describe the decay of the $\pi_2(1670)$ to $f_2(1270)\pi$
\begin{align}
    \mathcal{L} &= \cos\beta_t\left(g_c^{PT}\langle\pi_{2,\mu\nu}f_2^{\mu\nu}\pi\rangle + g_d^{PT}\langle\uppi_{2,\alpha\mu\nu}\mathfrak{f}_2^{\alpha\mu\nu}\pi\rangle\right),\label{lagpi2}
\end{align}
where $\uppi_{2,\alpha\mu\nu}=\partial_\alpha \pi_{2,\mu\nu}-\partial_\mu \pi_{2,\alpha\nu}$ , $\mathfrak{f}_2^{\alpha\mu\nu}=\partial^\alpha f_2^{\mu\nu}-\partial^\mu f_2^{\alpha\nu}$, and $\beta_t (=5.7^\circ)$ is the angle of mixing between the $2^{++}$ iso-singlets \cite{Zyla:2020zbs} (already included for later convenience). 
The experimental value of $D/S-$ratio for this decay is $-0.18\pm 0.06$ \cite{Zyla:2020zbs}.
The amplitude for this decay is
\begin{widetext}
\begin{align}
i\mathcal{M}&= i\cos\beta_t\left[g_c^{PT}\epsilon_{\mu\nu}(\vec{0},M_J)\epsilon^{\mu\nu\ast}(\vec{k},\lambda)\!+\! 2g_d^{PT}\left(k_0\cdot k_1\epsilon_{\mu\nu}(\vec{0},M_J)\epsilon^{\mu\nu\ast}(\vec{k_1},\lambda)\!-\!k_{0,\alpha} k_{1}^\nu\epsilon_{\mu\nu}(\vec{0},M_J)\epsilon^{\alpha\mu\ast}(\vec{k_1},\lambda)\right)\right]\label{ampJ2Ten}\\
&=\!i\cos\beta_t\Vast\{\begin{matrix*}[l]g_c^{PT}\dfrac{(M_{f_2}^2+2E_{f_2}^2)}{3M_{f_2}^2} +2 g_d^{PT} \dfrac{M_{\pi_2}}{M_{f_2}^2} E_{f_2} & M_J=\lambda=0\\g_c^{PT} \dfrac{E_{f_2}}{M_{f_2}} + g_d^{PT} \dfrac{M_{\pi_2} }{M_{f_2}}(k^2+2M_{f_2}^2) & M_J=\lambda=\pm1\\g_c^{PT}+2g_d^{PT}M_{\pi_2} E_{f_2}& M_J=\lambda=\pm 2 \end{matrix*}.\label{dertenamp}
\end{align}
\end{widetext}
For the $\pi_2(1670)\to f_2(1270)\pi$ decay, the allowed values of the relative angular momentum are $\ell=0,2,4$. Thus, from Eq. (\ref{amp1hel}), we get
\begin{align}
F^2_{20} &= \frac{1}{\sqrt{5}}G_0 + \sqrt{\frac{2}{7}}G_2+\frac{1}{\sqrt{70}}G_4\nonumber\\
F^2_{10} &= \frac{1}{\sqrt{5}}G_0 -\frac{1}{\sqrt{14}} G_2-\sqrt{\frac{8}{35}}G_4\nonumber\\
F^2_{00} &= \frac{1}{\sqrt{5}}G_0 -\sqrt{\frac{2}{7}} G_2+\sqrt{\frac{18}{35}}G_4,\label{helten}
\end{align}
where $G_0\,,\, G_2\,,\, \rm{and}\, G_4$ are the $\ell S$ coupling amplitudes for $\ell = 0\,,\, 2\,,\, 4$ respectively. The $G_\ell$'s can be calculated by solving the matrix equation
\begin{equation}
    \begin{pmatrix}
    \frac{1}{\sqrt{5}} & \sqrt{\frac{2}{7}} & \frac{1}{\sqrt{70}}\\
    \frac{1}{\sqrt{5}} & -\frac{1}{\sqrt{14}} & -\sqrt{\frac{8}{35}}\\
    \frac{1}{\sqrt{5}} & -\sqrt{\frac{2}{7}} & \sqrt{\frac{18}{35}}
    \end{pmatrix} \begin{pmatrix}
    G_0\\G_2\\G_4
    \end{pmatrix}= \begin{pmatrix}
    F^2_{20}\\F^2_{10}\\F^2_{00}
    \end{pmatrix}.\label{mateq}
\end{equation}
Solving for $G$'s, we get explicitly: 
\begin{widetext}
\begin{align}
  G_0&= \frac{1}{3
   \sqrt{5} M_{f_2}^2}\cos\beta_t\left[g_c^{PT} \left(2 E_{f_2}^2+6 E_{f_2} M_{f_2}+7 M_{f_2}^2\right)+g_d^{PT} M_{f_2} M_{\pi_2} \left(6E_{f_2}^2+18 E_{f_2} M_{f_2}+6
   M_{f_2}^2\right)\right]\label{G00}\\
   G_2 &= -\frac{1}{3 M_{f_2}^2}\sqrt{\frac{2}{7}}\cos\beta_t\left[g_c^{PT}
   \left(2 E_{f_2}^2+3 E_{f_2} M_{f_2}-5 M_{f_2}^2\right)+
   g_d^{PT} M_{f_2} M_{\pi_2}\left(3E_{f_2}^2-6 E_{f_2} M_{f_2}+3 M_{f_2}^2\right)\right]\label{G20}\\
   G_4 &= \frac{2}{M_{f_2}^2} \sqrt{\frac{2}{35}}\cos\beta_t \left[g_c^{PT} \left(E_{f_2}^2-2
   E_{f_2} M_{f_2}+M_{f_2}^2\right)- g_d^{PT} M_{\pi_2} M_{f_2} \left(2E_{f_2}^2-4 E_{f_2}
   M_{f_2}+2 M_{f_2}^2\right)\right].\label{G24}
\end{align}
\end{widetext}
The $D/S-$ratio can be used to estimate the ratio ($g_d^{PT}/g_c^{PT}$) of the coupling constants. We find that, in the absence of nonlocal interactions, $G_2/G_0=-0.018$, which is an order of magnitude smaller than the experimentally extracted value. Thus, nonlocal interactions become essential to explain the $D/S-$ratio for this decay. For the $D/S-$ratio to be equal to the value mentioned in the PDG, the ratio of the coupling constants must be $g_d^{PT}/g_c^{PT}=-0.209$ GeV$^{-2}$. This ratio is also of the same order of magnitude as $1/M_{\pi_2}^2$.\par
Finally, the decay width is given by
\begin{align}
    \Gamma_{\pi_2\to f_2\pi}&=f_{\pi_2f_2\pi}\frac{k\cos^2\beta_t}{40\pi M_{\pi_2}^2}\Bigg[(g_c^{PT})^2\left(\frac{4k^4}{9M_{f_2}^4}+\frac{10k^2}{3M_{f_2}^2}+5\right)\nonumber\\
    &+2(g_d^{PT})^2 M_{\pi_2}^2 M_{f_2}^2\left(\frac{k^4}{M_{f_2}^4}+\frac{10k^2}{M_{f_2}^2}+10\right)\nonumber\\
    &+ \frac{20}{3} g_c^{PT} g_d^{PT} E_{f_2} M_{\pi_2}\left(\frac{k^2}{M_{f_2}^2}+3\right)\Bigg],
\end{align}
where $f_{\pi_2f_2\pi}$ is the isospin symmetry factor. The contact and derivative interactions interfere destructively to give the above decay width, as evidenced by the fact that when $g_d^{PT}<0$, the last term is negative.
\begin{figure*}[t]
\centering
\includegraphics[scale=0.6]{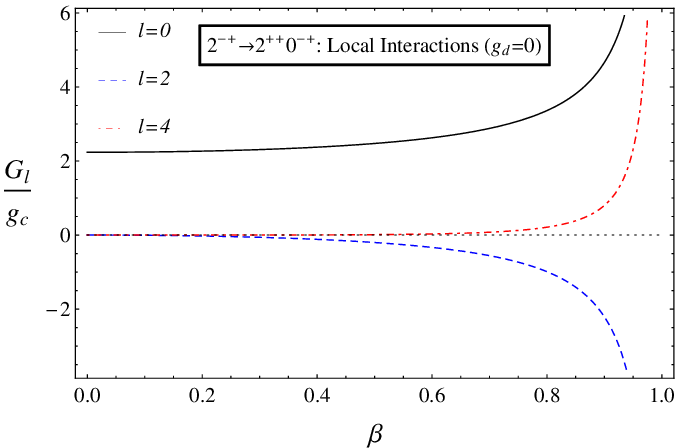}~\includegraphics[scale=0.65]{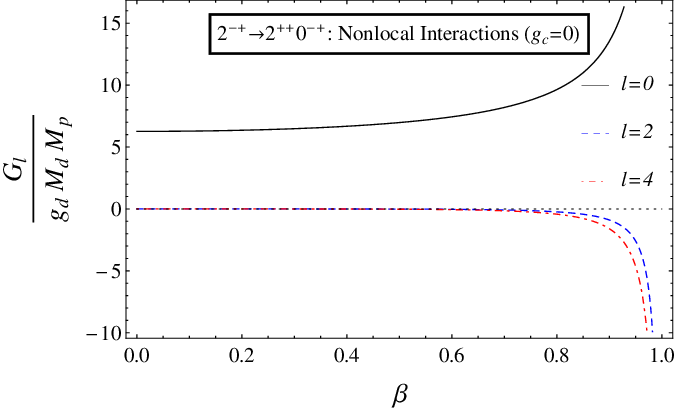}\\
\hspace*{-0.15in}\includegraphics[scale=0.63]{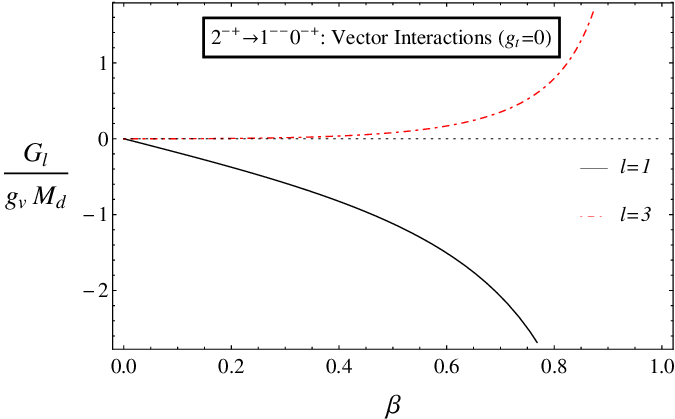}~~~~\includegraphics[scale=0.63]{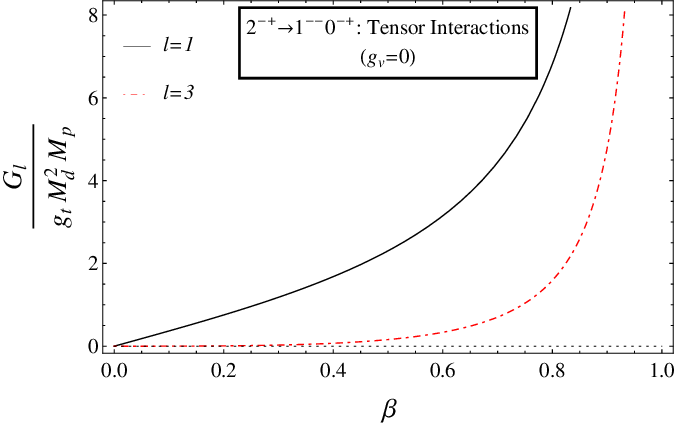}\\
\caption{Plots of the partial wave amplitudes (scaled appropriately) as functions of $\beta$: lower order terms (left), higher order terms (right); tensor decay mode (above), vector decay mode (right). See Sec (\ref{PWAanalysis}) for a detailed description.}\label{PWAxy}
\end{figure*}
\subsection{The \texorpdfstring{$\pi_2(1670)\to\rho\pi$}{} decay}\label{pwaderJ2}
The vector mode of decay is described by a dimension$-4$ operator that has a single derivative and generates ``vector" interactions, and a dimension$-6$ operator that has three derivatives and gives rise to ``tensor" interactions. The Lagrangian including these operators is
\begin{align}
\mathcal{L} &=ig_v^{PT}\langle\pi_{2,\mu\nu}\rho^\mu\partial^\nu\pi\rangle + ig_t^{PT}\langle\uppi_{2,\alpha\mu\nu}\uprho^{\alpha\mu}\partial^\nu\pi\rangle,\label{lagpi2vec}
\end{align}
where $g_v^{PT}$ and $g_t^{PT}$ are the respective coupling constants. The amplitude for the vector decay mode is
\begin{widetext}
\begin{align}
    i\mathcal{M} &= -g_v^{PT}~\epsilon_{\mu\nu}(\vec{0},M_J)\epsilon^{\mu\ast}(\vec{k_1},\lambda)k_2^\nu-g_t^{PT}~\Big[2k_0\cdot k_1\epsilon_{\mu\nu}(\vec{0},M_J)\epsilon^{\mu\ast}(\vec{k_1},\lambda)k_2^\nu - 2 k_{0,\mu}k_1^\alpha\epsilon_{\alpha\nu}(\vec{0},M_J)\epsilon^{\mu\ast}(\vec{k_1},\lambda)k_2^\nu\Big]\label{ampJ2Vec} \\
    \nonumber\\
    &= \frac{k}{\sqrt{2}} \Bigg\{\begin{matrix*}[l] (g_v^{PT} +2 g_t^{PT} M_{\pi_2} E_\rho) &\! M_J=\lambda=\pm 1\\
    \dfrac{2}{\sqrt{3}}\left(\dfrac{E_\rho}{M_\rho} g_v^{PT}+ 2 g_t^{PT} M_{\pi_2} M_\rho\right)&\! M_J=\lambda=0\end{matrix*}.
\end{align}
\end{widetext}
The helicity amplitudes can be derived using Eq. (\ref{amp2hel}) and Eq. (\ref{amp1LS}) just like the previous two cases. In this case, however, the allowed values of angular momentum are $\ell= 1,3$. The helicity amplitudes are related to the PWAs through the equations
\begin{align}
F^2_{10} &= \sqrt{\frac{3}{10}} G_1 + \frac{1}{\sqrt{5}}G_3\\
F^2_{00} &= \sqrt{\frac{2}{5}} G_1 - \sqrt{\frac{3}{5}} G_3.
\end{align}
Thus, we have two PWAs, $G_1$ and $G_3$, given by
\begin{widetext}
\begin{align}
  G_1&= -\sqrt{\frac{1}{15}}\frac{k}{M_\rho}\left[g_t^{PT} M_{\pi_2} M_\rho \left(4 M_\rho+6 E_\rho\right)+ g_v^{PT} (2 E_\rho+3M_\rho)\right]\label{G1}\\
  G_3&= \sqrt{\frac{2}{5}}\frac{k}{M_\rho}\left[g_t^{PT} M_{\pi_2} M_\rho \left(2 M_\rho - 2 E_\rho\right)+ g_v^{PT} (E_\rho-M_\rho)\right].\label{G3}
\end{align}
\end{widetext}
In order to reproduce the measured $F/P-$ratio ($-0.72\pm 0.16$, the coupling constants must have opposite sign: $g_t^{PT}/g_c^{PT} = -0.255$ GeV$^{-2}$, which is of the same order of magnitude as $1/M_{\pi_2}^2$. \par
The decay width is given by
\begin{align}
    \Gamma_{\pi_2 \to\rho\pi}&= f_{\pi_2\rho\pi}\frac{k}{40\pi M_{\pi_2}^2}\frac{k^2}{3}\Bigg[(g_v^{PT})^2\left(\frac{2k^2}{M_\rho^2}+5\right)\nonumber\\
    &\mkern-18mu+4(g_t^{PT})^2 M_\rho^2\left( \frac{3k^2}{M_\rho^2} +5\right)+ 20 g_v^{PT} g_t^{PT} E_\rho M_{\pi_2}\Bigg],\label{PTvecd}
\end{align}
where $f_{\pi_2\rho\pi}$ is the isospin symmetry factor. Again, destructive interference between the different interaction types takes place. 
\begin{figure*}[ht]
\centering
\includegraphics[scale=0.475]{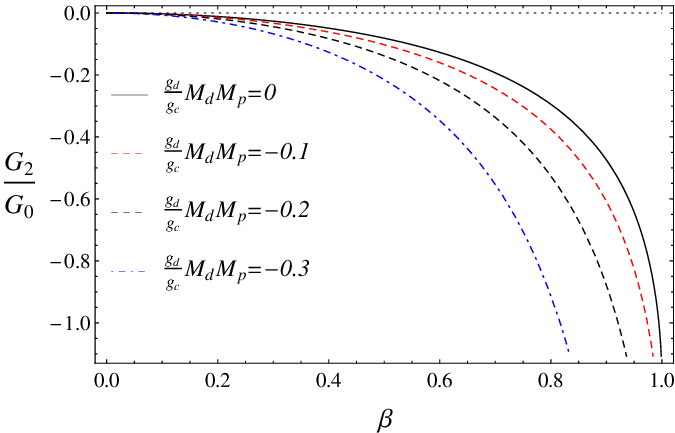}~\includegraphics[scale=0.475]{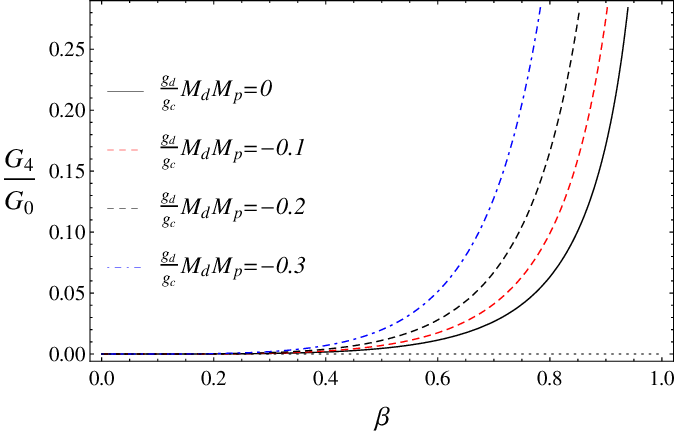}~\includegraphics[scale=0.475]{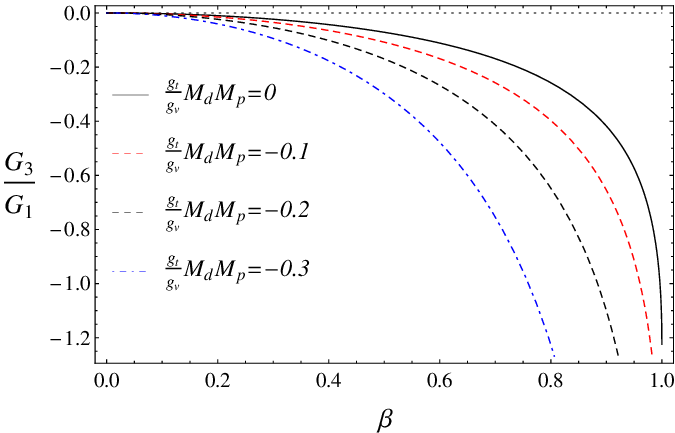}
\caption{The $D/S-$, $G/S-$, and the $F/P-$ratios (left, center, and right respectively) as functions of $\beta$ for representative values of $\dfrac{g_d}{g_c} M_p M_{d,1}$ and $\dfrac{g_t}{g_v} M_p M_{d,1}$.}\label{PWAratxy}
\end{figure*}
\subsection{Analysis of the PWAs}\label{PWAanalysis}
We now look at the partial wave amplitudes $G_0$ - $G_4$. To study the behavior of the PWAs, we look at the $\ell=0,2,4$ amplitudes mentioned in Eq. (\ref{G20}) - Eq. (\ref{G24}) and the $\ell=1,3$ amplitudes given in Eq. (\ref{G1}) and Eq. (\ref{G3}). Below, we rewrite these equations in terms of the Lorentz factor\footnote{All the figures discussed in this subsection are plotted as functions of $\beta$, which is related to $\gamma$ as $\gamma=\frac{E_{d,1}}{M_{d,1}} = \frac{1}{\sqrt{1-\beta^2}}$. This is because the range of $\beta$ is $[0,1]$, whereas that of $\gamma$ is $[1,\infty)$.} $(\gamma)$. In all these expressions, we have used the symbols $M_p$, and $M_{d,1}$ to denote the masses of the decaying (parent, $p$) state and the heavier decay product (vector/tensor meson, daughter, $d,1$) respectively.
\begin{widetext}
\begin{align}
 G_0&= \frac{1}{3
   \sqrt{5}}\left[g_c \left(2\gamma^2+6 \gamma+7\right)+g_d M_{d,1} M_p \left(6\gamma^2+18 \gamma+6\right)\right]\label{G00g}\\
   G_2 &= -\frac{1}{3 }\sqrt{\frac{2}{7}}\left[g_c
   \left(2 \gamma^2+3 \gamma -5 \right)+
   g_d M_{d,1}  M_p\left(3\gamma^2-6 \gamma +3 \right)\right]\label{G20g}\\
   G_4 &= 2 \sqrt{\frac{2}{35}} \left[g_c \left(\gamma^2-2
   \gamma + 1\right)- g_d M_{d,1} M_p  \left(2\gamma^2-4 \gamma
   +2 \right)\right]\label{G24g}\\
 G_1&= -\sqrt{\frac{1}{15}}\sqrt{\gamma^2-1}M_{d,1}\left[g_v (2 \gamma+3) + g_t M_p M_{d,1} \left(4 +6\gamma\right)\right]\label{G1g}\\
  G_3&= \sqrt{\frac{2}{5}}\sqrt{\gamma^2-1}M_{d,1}\left[g_v (\gamma-1) + g_t M_p M_{d,1} \left(2\gamma - 2 \right)\right]\label{G3g}
\end{align}
\end{widetext}
These expressions are valid for the decay of any $J=2$ state to any $J=2$ or $J=1$ state, irrespective of their charge conjugation quantum number or the states being ground states or excited states. For example, the $2^{--}\to 1^{+-}0^{-+}$ decays proceed with $\ell=1,3$, and hence, the corresponding PWAs will be given by Eq. (\ref{G1g}) and Eq. (\ref{G3g}). Similarly, the $2^{--}\to 2^{++}0^{-+}$ decays proceed with $\ell=0,2,4$ and their amplitudes will be given by Eq. (\ref{G00g}) - (\ref{G24g}). The difference between these decays and the ones studied in the present work lies in the value of the coupling constants. The following observations are in order:
\begin{enumerate}
\item In the absence of the derivative/tensor interactions, the PWAs depend only on the Lorentz factor. 
\item The amplitudes mentioned in Eq. (\ref{G00g}) - Eq. (\ref{G1g}) are plotted in Fig. (\ref{PWAxy}). The plots on the top row show the contributions of the contact and derivative interactions to the PWAs as functions of $\beta$. On the bottom row are the corresponding vector and tensor contributions to the vector mode of the pseudotensor decay. All the higher partial waves ($G_1,~\!\!G_2,~\!\!G_3,\text{ and } G_4$) vanish as the momentum carried by the decay products goes to zero (i.e, nonrelativistic limit). In this limit, the $S-$ wave has the amplitude proportional to $\sqrt{5}(g_c+2g_d M_p M_{d,1})$. We infer from Eq. (\ref{G1g}) that the $P-$wave amplitude also vanishes in the nonrelativistic limit, due to an overall multiplying factor of $\sqrt{\gamma^2-1}$. 
\item In the ultrarelativistic limit (i.e, $\beta\to1$), the higher partial waves dominate over the $S-$wave and the $P-$wave. In this case, the $D/S$, $G/S$, and the $F/P-$ratios become much larger than 1, as can be seen from Fig. (\ref{PWAratxy}). In the tensor mode of the decay of the pseudotensor meson, the $G/S-$ratio becomes larger than the $D/S-$ratio. The behavior of the PWAs in this region is dominated by the derivative/tensor interaction (i.e, higher order contributions to the Lagrangian).
\end{enumerate}

\section{Coupling constants, isoscalar mixing angles, and their consequences }\label{RnDPWA}
In this section, we employ the formalism described in the previous section in order to evaluate PWA for various nonet memebrs, to detemrine the coupling constants, the strange-nonstrange mixing angle of the isoscalar members of a given nonet, and to discuss their consequences.

\subsection{\texorpdfstring{$J^{PC}=1^{++}$}{JPC=1++}}
In this subsection, we demonstrate the working of our model by applying it to the $J^{PC}=1^{++}$ nonet. In this sector, the Lagrangian has three parameters: the coupling constants $g_c^A$ and $g_d^A$, and the isoscalar mixing angle\footnote{\bf Here, and everywhere else, the Lagrangian describing the decays of the isoscalars is identical to that for the decays of the isovectors, except for the isospin symmetry factors.} $\theta_a$. The mixing angle enters the Lagrangian through the scheme
\begin{align}
    \begin{pmatrix}
    |f_1\rangle\\
    |f_1^\prime\rangle
    \end{pmatrix}&= \begin{pmatrix}
    \cos\theta_a & \sin\theta_a\\-\sin\theta_a & \cos\theta_a
    \end{pmatrix}
    \begin{pmatrix}
    |\bar{n}n\rangle_a\\
    |\bar{s}s\rangle_a
    \end{pmatrix},
\end{align}\label{avmix}
where $|\bar{s}s\rangle_a$ and $|\bar{n}n\rangle_a~ (= \frac{1}{\sqrt{2}}(|\bar{u}u\rangle_a + |\bar{d}d\rangle_a))$, where the subscript $`a'$ represent axial-vector states, are the strange and non-strange iso-singlet states respectively. We also have three data points: the $D/S-$ratio and the width of the $a_1(1260)\to\rho\pi$ decay, and the width of the $f_1'(1420)\to K^*K$ decay. The PDG lists the $D/S-$ratio for the $a_1(1260)\to\rho\pi$ decay as $-0.062\pm 0.02$ \cite{Zyla:2020zbs}. Since the $\rho\pi$ channel is the dominant channel for the decay of the $a_1(1260)$, we take the total width ($420\pm35$ MeV) of the $a_1(1260)$ as the width of this channel. The width of the $f_1'(1420)\to K^*K$ decay can be estimated as $44.5\pm 4.2$ MeV, using the branching fraction listed in the PDG \cite{Zyla:2020zbs}. Since the number of unknowns is the same as the number of data points available, the values of the parameters can be estimated without resorting to a statistical fit. We, however, define the $\chi^2$ function so as to calculate the errors in the values of the parameters and those in the widths and PWA ratios. The input values are listed in Table \ref{1plusip} and the values of the parameters so obtained are listed in Table \ref{1pluspar}.\par
\begin{table}[ht]
    \centering
    {\renewcommand{\arraystretch}{1.5}
    \begin{tabular}{|c|c|c|}
    \hline
    \rule{0pt}{2em} Decay & Width (MeV) & $D/S$ \cite{Zyla:2020zbs}\\\hline
    $a_1(1260)\to\rho\pi$ & $420\pm35$ & $-0.062\pm0.02$\\
    $f_1'(1420)\to K^*K$ & $44.5\pm4.5$ & $---$\\\hline
    \end{tabular}}
    \caption{Input values used to extract the values listed in Table \ref{1pluspar}.}
    \label{1plusip}
\end{table}
We use the values of the parameters thus obtained to estimate the $D/S-$ratios and the widths for the kaonic decays of $f_1(1285)$ and $f_1'(1420)$. These values are listed in Table \ref{1plusds}. Of these two decays, the $f_1(1285)\to K^*K$ is decay is sub-threshold and hence kinematically suppressed. We perform a spectral integration over the final $K^*$ to obtain the width and the $D/S-$ratio for this decay (see Appendix \ref{specintappe} for details). We make the following observations:
\begin{table}[ht]
    \centering
    {\renewcommand{\arraystretch}{1.5}
    \begin{tabular}{ccc}
    \hline\hline
        $g_c^A$ (GeV) & $g_d^A$ (GeV$^{-1}$) & $\theta_a$\\\hline
        $3.89\pm 0.75$ & $-0.32\pm0.37$ & $(24.9\pm 3.2)^\circ$\\\hline
    \end{tabular}}
    \caption{Values of the parameters used in the decays of the axial-vector mesons.}
    \label{1pluspar}
\end{table}
\begin{enumerate}
	\item The small value of the $D/S-$ratio for the $1^{++}$ decays indicate that the $D-$wave interactions, which are predominantly derivative interactions, play only a minor role. Correspondingly, the coupling constant $g_d^A$ has a small value (compatible with zero). In other words, $g_c^a/g_d^A \ll M_{a_1}^2$.
	\item The mixing of the isoscalars is an important feature of QCD.
	The value of the $1^{++}$ isoscalar mixing angle obtained in the present work ($\theta_a=(24.9\pm3.2)^\circ$) is consistent with the experimental value ($\pm(24.0^{+3.7}_{-3.4})^\circ$) reported in \cite{LHCb:2013ged} as well as the lattice results ($(31\pm2)^\circ$) \cite{Dudek:2011tt} (see also Refs. \cite{Jiang:2020eml,Liu:2014doa} for comparison). The iso-singlet mixing angles in the $J=1$ sector are sensitive to the masses and mixing angle of the corresponding kaons, if viewed through the Gell-Mann-Okubo (GMO) mass relations. However, the $\bar{B}^0\to J/\psi f_1(1285)$ and the $\bar{B}^0_s\to J/\psi f_1(1285)$ decays provide a much cleaner view into the mixing between $f_1(1285)$ and $f_1^\prime(1420)$. The ratio of the branching fractions of these two decays is proportional to $\tan^2\theta_a$ and, more importantly, independent of the kaonic mixing angle \cite{LHCb:2013ged,Stone:2013eaa}. However, we note that, this measurement cannot give us the information regarding the sign of the mixing angle.
\begin{table}[ht]
    \centering
    {\renewcommand{\arraystretch}{1.5}
    \begin{tabular}{|c|c|c|}
    \hline
    \multicolumn{3}{|c|}{Wdith (MeV)}\\\hline
    \rule{0pt}{2em} Decay & Theory & PDG \cite{Zyla:2020zbs} \\\hline
    $f_1(1285)\to K^*K$ & $4.78\pm0.57$& not seen\\\hline
    \multicolumn{3}{|c|}{$D/S-$ratio}\\\hline
    \rule{0pt}{2em} Decay & Theory & PDG \cite{Zyla:2020zbs} \\\hline
     $f_1(1285)\to K^*K$ & $-(0.436\pm 0.87)\times 10^{-3}$ & $---$\\
    $f_1'(1420)\to K^*K$ & $-0.0116\pm 0.005$ & $---$\\\hline
    \end{tabular}}
    \caption{Predictions based on the parameters listed in Table \ref{1pluspar}. See text for details of the calculations.}
    \label{1plusds}
\end{table}
	\item We compare the values of the $D/S-$ratios we obtain with those extracted using the $^3P_0$ model \cite{Barnes:1996ff}. A brief review of the $^3P_0$ model is presented in Appendix \ref{3P0rev} and the corresponding results are listed in Table \ref{com3P0}. We see that, for the $a_1(1260)\to\rho\pi$ decay, our value agrees with PDG average \cite{Zyla:2020zbs}, where as the $^3P_0$ values are compatible with the values obtained by the E852 \cite{Chung:2002pu}, OPAL \cite{OPAL:1997was}, and the ARGUS \cite{ARGUS:1992olh} collaborations. Our value is nearly $2.5$ times smaller than the $^3P_0$ one. This carries over to the decay of the isoscalar meson as well. Our estimate for the $D/S-$ratio of the $f_1'(1420)\to K^*K$ decay is nearly $3.5$ times smaller than that from the $^3P_0$ model.
\end{enumerate}

\subsection{\texorpdfstring{$J^{PC}=1^{+-}$}{JPC=1+-}}
    \begin{table}[ht]
    \centering
    {\renewcommand{\arraystretch}{1.5}
    \begin{tabular}{|c|c|c|}
    \hline
    \rule{0pt}{2em} Decay & Width (MeV) & $D/S$ \cite{Zyla:2020zbs}\\\hline
    $b_1(1235)\to\omega\pi$ & $110\pm7$\cite{Divotgey:2013jba} & $0.277\pm0.027$\\
    $h_1'(1415)\to K^*K$ & $90\pm15$\cite{Zyla:2020zbs} & $---$\\\hline
    \end{tabular}}
    \caption{Input values used to extract the values listed in Table \ref{pvpar}.}\label{pvip}
\end{table}
We now turn our attention to the decays of the pseudovector mesons. The Lagrangian describing the $1^{+-}\to 1^{--}0^{-+}$ decays is similar to the one written in Eq. (\ref{lagJ1}). Thus, we have three parameters: the coupling constants $g_c^B$ and $g_d^B$, and the isoscalar mixing angle $\theta_{pv}$. Similar to the case of axial-vectors, the mixing angle is defined through the relation
\begin{align}
    \begin{pmatrix}
    |h_1\rangle\\
    |h_1^\prime\rangle
    \end{pmatrix}&= \begin{pmatrix}
    \cos\theta_{pv} & \sin\theta_{pv}\\-\sin\theta_{pv} & \cos\theta_{pv}
    \end{pmatrix}
    \begin{pmatrix}
    |\bar{n}n\rangle_{pv}\\
    |\bar{s}s\rangle_{pv}
    \end{pmatrix},
\end{align}\label{pvmix}
where the subscript $`pv'$ implies psuedovector. The values of these three parameters can be obtained using the width and $D/S-$ratio of the $b_1(1235)\to\omega\pi$ decay, and the width of the $h_1'(1415)\to K^*K$ decay. The values of the input parameters are listed in Table \ref{pvip}. We note that, the PDG does not list the partial widths of the decays of the $1^{+-}$ mesons. Hence, we have used the values obtained in an earlier work \cite{Divotgey:2013jba} for the width of the $b_1(1235)\to\omega\pi$ decay, and the total width of the $h_1'(1415)$ as the width of the $h_1'(1415)\to K^*K$ decay, as this is the only observed channel \cite{Zyla:2020zbs}. The values of the parameters obtained using these data and the associated errors are listed in Table \ref{pvpar}. 
 \begin{table}[ht]
    \centering
    {\renewcommand{\arraystretch}{1.5}
    \begin{tabular}{ccc}
    \hline\hline
        $g_c^B$ (GeV) & $g_d^B$ (GeV$^{-1}$) & $\theta_{pv}$\\\hline
        $6.36\pm 0.72$ & $-4.37\pm0.37$ & $(25.2\pm 3.1)^\circ$\\\hline
    \end{tabular}}
    \caption{Values of the parameters used in the decays of the pseudovector mesons.}
    \label{pvpar}
\end{table}
\begin{enumerate}
    \item In the decays of the pseudovector mesons, the $D-$waves interfere largely constructively with the $S-$waves. It should be noted that there exists a small phase difference of $(10\pm 5)^\circ$ between the $D-$wave and the $S-$wave in the $b_1(1235)\to\omega\pi$ decay \cite{Zyla:2020zbs}. However, we have not been able to reproduce this phase difference. Further, in the absence of the derivative interactions, the $D/S-$ratio of the $b_1(1235)\to\omega\pi$ decay is negative and is nearly equal to the corresponding ratio for the $a_1(1260)\to\rho\pi$ decay. The nonlocal interactions introduced in the form of the dimension$-5$ operators contribute a large amount to the $D/S-$ratio to make it significantly large and positive. This signifies that the nonlocal interactions play a crucial role in the pseudovector sector.
    \item The values of the parameters listed in Table \ref{pvpar} are significantly different from the values derived in Ref. \cite{Divotgey:2013jba}. This can be attributed to two reasons: the introduction of the derivative interactions in the pseudovector sector, and the mixing of the iso-singlet states. Derivative interactions were used to analyse the decay of $b_1(1235)$ in \cite{Jeong:2018exh}. The coupling constants $g_1$ and $g_2$ mentioned there have been rendered dimensionless by the multiplying/dividing mass term. The ratio of the two coupling constants used in our work, $g_d^B/g_c^B$ matches the corresponding value from Ref. \cite{Jeong:2018exh}\footnote{We point out a misprint in \cite{Jeong:2018exh}, i.e, the value of $g_1$ must be $-8.37$ instead of the $-1.34$ given in the article.}, as can be seen from Table \ref{1pluspar}. The absolute values of the coupling constants are slightly different as we have included the iso-singlet states as well in our work as opposed to only the iso-triplet in \cite{Jeong:2018exh}. However, as far as the $D/S-$ratio is concerned, only the ratio of the coupling constants matters. Further, we have taken the partial decay width for the $b_1(1235)\to\omega\pi$ decay as $110\pm 7$ MeV, instead of the full width of $142\pm 9$ MeV.
    \item The value of the coupling constant $g_c^B$ is nearly twice that of $g_c^A$, as shown in Table \ref{1pluspar}. This is particularly interesting, as the pseudovector states differ from the axial-vector states only in the charge conjugation, the decay products belong to the same set of nonets, and the 3-momenta carried by the decay products in both the cases are nearly the same.
    \item We also note that, in general, the influence of the $D-$wave on the decay of the mesons reduces as the 3-momenta of the decay products decreases, as seen by the decreasing values of the $D/S-$ratio. This is a feature we observe irrespective of the spin of the decaying state. This indicates that, when a meson decays into a closely lying state (specifically, if the associated 3-momentum is small) the angular distribution of the decay products is mostly spherical, and one may not lose much information if the higher partial waves are not included while analysing the experimental data.
    \begin{table}[ht]
    \centering
    {\renewcommand{\arraystretch}{1.5}
    \begin{tabular}{|c|c|c|}
    \hline
     \multicolumn{3}{|c|}{Wdith (MeV)}\\\hline
    \rule{0pt}{2em} Decay & Theory & PDG \cite{Zyla:2020zbs} \\\hline
    $h_1(1170)\to\rho\pi$ & $146\pm 14$ & seen\\\hline
    \multicolumn{3}{|c|}{$D/S-$ratio}\\\hline
    \rule{0pt}{2em} Decay & Theory & PDG \cite{Zyla:2020zbs} \\\hline
    $h_1(1170)\to\rho\pi$ & $0.281\pm 0.035$ & $---$\\
   $h_1'(1415)\to K^*K$ & $0.021\pm 0.001$ & $---$\\\hline
    \end{tabular}}
    \caption{Predictions based on the parameters listed in Table \ref{pvpar}. See text for details of the calculations.}
    \label{pvds}
\end{table}
    \item The mixing angle between the pseudovector isoscalars comes out to be larger than the value extracted by the BESIII collaboration \cite{Ablikim:2018ctf}. Our estimate of the mixing angle is $(25.2\pm 3.1)^\circ$, where as the BESIII collaboration reports a nearly zero mixing among the strange and non-strange states ($\theta_{pv} = (0.6\pm2.6)^\circ$), which agrees with the lattice results ($(3\pm 1)^\circ$) \cite{Dudek:2011tt}. One should however note that, the analysis of the BESIII is based on the mass of the $h_1(1415)$ and is very much sensitive to the value of the kaonic mixing angle which, in turn, is based on the GMO mass relations \cite{Cheng:2011pb}. Thus, a better avenue, similar to the case of the axial-vector, is needed to get a good insight into the mixing of pseudovector isosinglets. According to our analysis, a significantly large mixing angle is necessary to explain the smaller width of $h_1(1415)$. With the mixing angle we obtain, the $h_1(1415)$ can be seen as a mixture of approximately $83\%$ $|\bar{s}s\rangle$ and $17\%$ $|\bar{n}n\rangle$ and vice versa for the $h_1(1170)$.
    \begin{table}[ht]
    \centering
    {\renewcommand{\arraystretch}{1.5}
    \begin{tabular}{cccc}
    \hline\hline
        $g_c^{PT}$ (GeV) & $g_d^{PT}$ (GeV$^{-1}$) & $g_v^{PT}$ & $g_t^{PT}$ (GeV$^{-2}$)\\\hline
        $39\pm 13$ & $-8.16\pm 2.95$ & $-9.44\pm 1.24$ & $2.41\pm 0.50$\\\hline
    \end{tabular}}
    \caption{Values of the parameters used in the decays of the pseudotensor mesons.}\label{2pluspar}
\end{table}
 \begin{table*}[ht]
    \centering
    {\renewcommand{\arraystretch}{1.5}
    \begin{tabular}{|c|c|c|c|}
    \hline
    \rule{0pt}{2em} Decay & Width (MeV) & $D/S$ \cite{Zyla:2020zbs} & $F/P$ \cite{Zyla:2020zbs}\\\hline
    $\pi_2(1670)\to f_2(1270)\pi$ & $146.4\pm 9.7$ & $-0.18\pm 0.06$ & $\times\times\times$\\
    $\pi_2(1670)\to\rho\pi$ & $80.6\pm 10.8$ & $\times\times\times$ & $-0.72\pm 0.16$\\\hline
    \end{tabular}}
    \caption{Input values used to extract the values listed in Table \ref{2pluspar}.}\label{ptip}
\end{table*}
 \begin{table*}[ht]
    \centering
    {\renewcommand{\arraystretch}{1.5}
    \begin{tabular}{|c|c|c|c|c|}
    \hline
    \rule{0pt}{2em} Decay & Width (MeV)& $D/S$ & $G/S$ & $F/P$\\\hline
    $\pi_2(1670)\to f_2(1270)\pi$ & \textbf{Input} & \textbf{Input} & $0.0042\pm 0.0014$ & $\times\times\times$\\
    $\pi_2(1670)\to f_2'(1520)\pi$ & $0.43\pm 0.21$ & $0.00925\pm 0.0031$ & $-(7.49\pm2.7)\times 10^{-6}$ & $\times\times\times$\\
    $\pi_2(1670)\to K^*K$ & $5.11 \pm 1.4$ & $\times\times\times$ & $\times\times\times$ & $-0.447\pm 0.099$\\\hline
    \end{tabular}}
    \caption{Predictions based on the parameters listed in Table \ref{2pluspar}. See text for details of the calculations.}
    \label{ptdsfp}
\end{table*}
    \item The parameters obtained have been used to calculate the $D/S-$ratio and the width of the $h_1(1170)\to\rho\pi$ decay as well as the $D/S-$ratio of the $h_1'(1415)\to K^*K$ decay. These values are listed in Table \ref{pvds}. We observe that the $D/S-$ratio for the $h_1(1170)\to\rho\pi$ decay is marginally higher than that for the $b_1(1235)\to\omega\pi$ decay even though the $\omega\pi$ carry significantly larger 3-momentum ($348$ MeV) than the $\rho\pi$ ($303$ MeV). This is because, the $S-$wave amplitude in the $b_1(1235)\to\omega\pi$ decay is nearly $36\%$ higher than that of the $h_1(1170)\rho\pi$ decay whereas the $D-$wave amplitude is only $\sim 33\%$ larger.
    \begin{figure*}[t]
        \centering
        \includegraphics[scale=0.7]{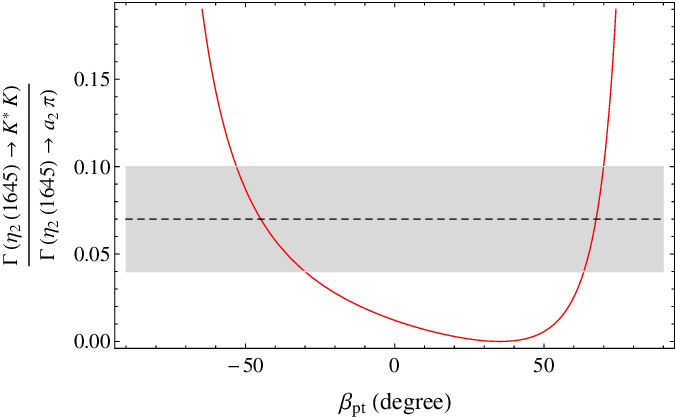}~\includegraphics[scale=0.7]{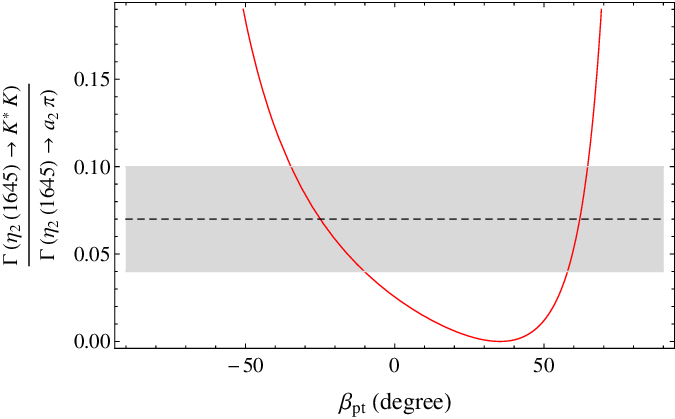}
        \caption{Plot of ratio of the partial widths of the two modes of decays of $\eta_2(1645)$ discussed in the text, as a function of the mixing angle: (left) including higher order terms and (right) without the higher order terms. The shaded region represents the uncertainty in the experimental value.}
        \label{eta2pw}
    \end{figure*}
    \item We compare our estimates of the $D/S-$ratios of the decays of the pseudovector mesons with those obtained from the $^3P_0$ model. Unlike the values for the decays of the axial-vector mesons, our values are nearly in good agreement with those from the $^3P_0$ model  (see Appendix B and the Table \ref{com3P0} therein).
\end{enumerate}

    \begin{table*}[ht]
    \centering
    {\renewcommand{\arraystretch}{1.5}
    \begin{tabular}{|c|c|c|}
    \hline
    Decay & \multicolumn{2}{c|}{Width (MeV)} \\\cline{2-3}
    & $\beta_{pt}=-(44.2^{+11}_{-15})^\circ$ & $\beta_{pt}+(67.3^{+2.5}_{-4.1})^\circ$\\\hline
    $\eta_2(1645)\to a_2\pi$ & $185\pm 12$ & $54.1\pm 3.6$\\
    $\eta_2(1645)\to K^\ast K$ & $12.9\pm 3.28$ & $3.78\pm 0.95$\\\hline
    $\eta_2(1870)\to a_2\pi$ & $50.7\pm 9.2$ & $86.9\pm 23$\\
    $\eta_2(1870)\to f_2(1270)\eta$ & $0.16\pm0.84$ & $0.56\pm2.1$\\
    $\eta_2(1870)\to K^\ast K$ & $0.65\pm0.15$ & $15.83\pm3.2$\\\hline
	\end{tabular}}
	\caption{Widths of the of decays of $\eta_2(1645)$ and $\eta_2(1870)$ studied in this work. The uncertainties are from the uncertainties in the coupling constants. The uncertainties in the mixing angle have not been considered.}\label{etatab}
\end{table*} 

\subsection{\texorpdfstring{$J^{PC}=2^{-+}$}{JPC=2-+}}
\begin{table*}[ht]
    \centering
    {\renewcommand{\arraystretch}{1.5}
    \begin{tabular}{|c|c|c|c|}
    \hline
        Decay & $D/S$ & $G/S$ & $F/P$ \\\hline
        $\eta_2(1645)\to a_2(1320)\pi$ & $-0.089 \pm 0.029$ & $0.0011\pm 0.0004$ & $--$ \\
        $\eta_2(1645)\to K^\ast K$ & $--$ & $--$ & $-0.32\pm 0.07$ \\\hline
    \end{tabular}}
    \caption{The ratios of the PWAs for the decays of $\eta_2(1645)$ discussed in the text.}
    \label{eta2tabrat}
\end{table*}
We now turn our attention to the decay of the $2^{-+}$ mesons. Here, we have analyzed two kinds of decays: $2^{-+}\to 2^{++}0^{-+}$ (tensor mode) and $2^{-+}\to 1^{--}0^{-+}$ (vector mode). The tensor decay mode is described by the Lagrangian given in Eq. (\ref{lagpi2}) and the vector mode by Eq. (\ref{lagpi2vec}). Each of the Lagrangians contain two parameters: the tensor mode coupling constants $g_c^{PT}$ and $g_d^{PT}$, and the vector mode coupling constants $g_v^{PT}$ and $g_t^{PT}$, the values of which are listed in Table \ref{2pluspar}. The data used as inputs to derive the values of these parameters are listed in Table \ref{ptip}. The decay widths listed in Table \ref{ptip} have been calculated using the branching fractions listed in the PDG \cite{Zyla:2020zbs}. The values of the parameters can be estimated similar to the case of the $J^P=1^+$ nonets. Using these parameters, we calculate the ratios of the PWAs and the widths for the $\pi_2(1670)\to f_2'(1520)\pi$, and $\pi_2(1670)\to K^*K$ decays. These values are listed in Table \ref{ptdsfp}.
We make the following observations:
\begin{enumerate}
    \item In the absence of derivative interactions, the $D/S-$ratio for the $\pi_2(1670)\to f_2(1270)\pi$ decay is an order of magnitude smaller than the experimentally extracted value. In the case of the $\pi_2(1670)\to\rho\pi$ decay, in the absence of the tensor interactions, the value of the $F/P-$ratio comes out to be less than $1/5^{th}$ of the experimental value. Thus, the nonlocal/tensor interactions contribute to a large extent to the decay of the tensor mesons. A closer inspection of the amplitudes of the individual partial waves (Eq. (\ref{G20})$-$(\ref{G24})) show that the coupling constant for the derivative interactions decides the sign of each amplitude in case of the tensor decay mode.
    \item According to our analysis, the contributions of the $\ell=4$ wave to the decay of tensor mesons is nearly two orders of magnitude smaller than the $\ell=2$ waves. But, the $D-$waves and the $G-$waves interfere destructively. Here again, the nonlocal interactions play an important role in deciding the phases of these waves relative to the $S-$wave.
    
    \begin{figure*}[t]
        \centering
        \includegraphics[scale=0.7]{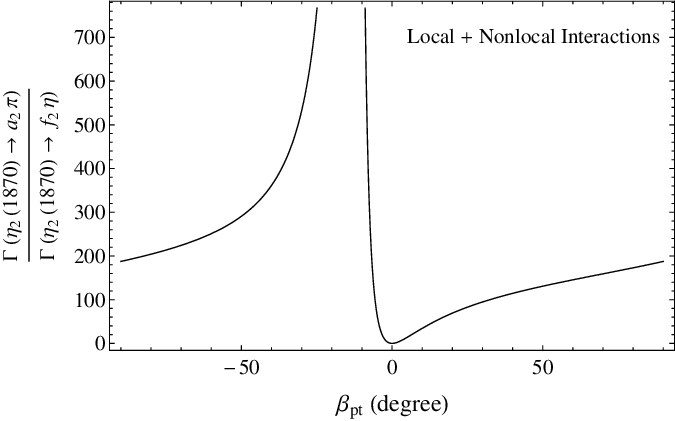}~\includegraphics[scale=0.7]{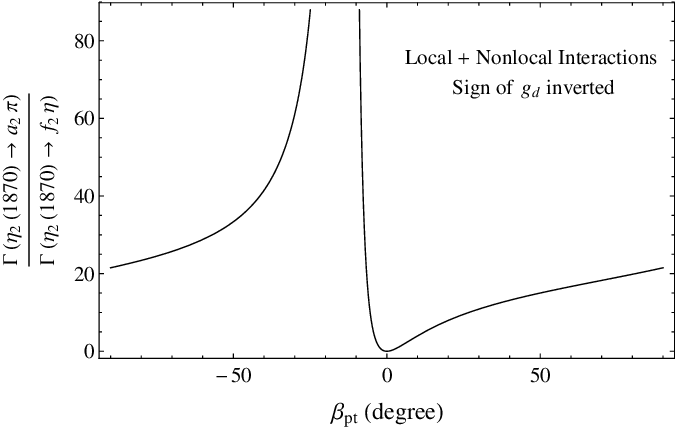}\\
         \includegraphics[scale=0.7]{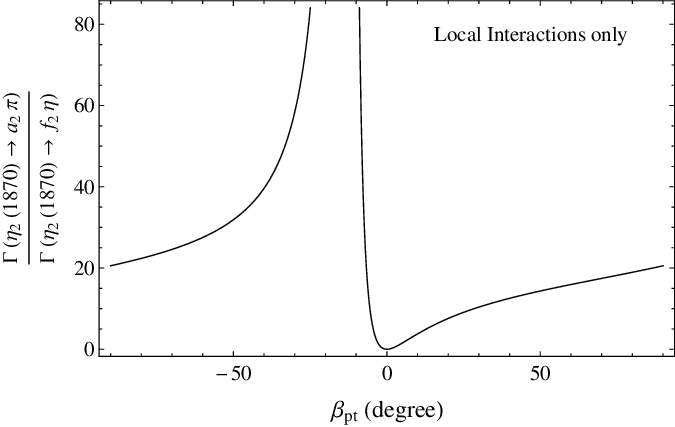}~\includegraphics[scale=0.7]{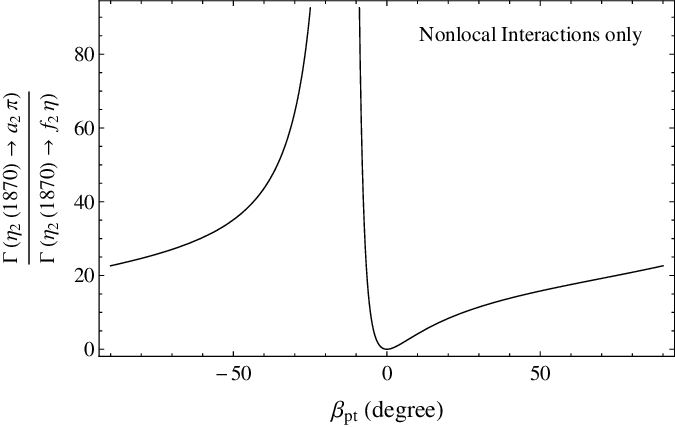}
        \caption{Plot of the ratio partial widths of the two modes of decays of $\eta_2(1870)$ discussed in the text: (above, left) as a function of mixing angle; (above, right) with the sign of the coupling constant for derivative interaction flipped; (below) contributions of the local (left) and nonlocal interactions (right) as a function of mixing angle.}
        \label{eta2hrat}
    \end{figure*}
    
    \item The $J=2$ isosinglets pose a special problem. The nature of $\eta_2(1870)$ is still a mystery. The absence of evidence for the $K^\ast K$ decay mode makes it difficult to interpret it as the heavier sibling of the $\eta_2(1645)$ \cite{Anisovich:2010nh,Klempt:2007cp}. Without this state, though, the $2^{-+}$ nonet is incomplete. Further, the angle of mixing between the two iso-singlet is still an open problem. The mixing angle ($\beta_{pt}$), given by the scheme
    \begin{align}
    \quad\quad\begin{pmatrix}
    |\eta_2\rangle\\
    |\eta_2^\prime\rangle
    \end{pmatrix}&= \begin{pmatrix}
    \cos\beta_{pt} & \sin\beta_{pt}\\-\sin\beta_{pt} & \cos\beta_{pt}
    \end{pmatrix}
    \begin{pmatrix}
    |\bar{n}n\rangle_{pt}\\
    |\bar{s}s\rangle_{pt}
    \end{pmatrix}
    \end{align}
     where `$pt$' stands for pseudotensor, is expected to be large in this sector as the $2^{-+}$ mesons are heterochiral states \cite{Giacosa:2017pos}. A recent work \cite{Koenigstein:2016tjw} reported that the mixing angle for the $\eta_2(1645)$ and $\eta_2(1870)$ to has to be larger than the value ($14.8^\circ$) derived using the GMO relations to properly fit the decay widths. The value of the mixing angle was found to be $-42^\circ$\cite{Koenigstein:2016tjw}. However, this failed to reproduce the ratio of branching fractions of the decays $\eta_2(1870)\to a_2\pi$ to $\eta_2(1870)\to f_2\eta$. The value of this ratio calculated in Ref. \cite{Koenigstein:2016tjw} was $23.5$, which is an order larger than the value $1.7\pm0.4$ accepted by PDG \cite{Zyla:2020zbs} (but, close to the value extracted by the WA102 collaboration \cite{WA102:1999ybu}). For the present analysis, we proceed assuming that the $\eta_2(1870)$ is the iso-singlet of the pseudotensor nonet. The Lagrangian that describes the decays of these isoscalars are similar to the ones given in Eq. (\ref{lagpi2}) and Eq. (\ref{lagpi2vec}), except for the mixing and the isospin factors. In Fig. (\ref{eta2pw}), we have plotted the ratio of the widths of the $\eta_2(1645)\to a_2(1320)\pi$ and $\eta_2(1645)\to K^\ast K$ decays as functions of the mixing angle. The dashed horizontal line in the Fig. (\ref{eta2pw}) represents the experimental value of this ratio ($0.07\pm0.03$) \cite{WA102:1997gkz,Zyla:2020zbs}. We see that two values of mixing angle can reproduce this data: $\beta_{pt}=-(44.2^{+11}_{-15})^\circ$ and $+(67.3^{+2.5}_{-4.1})^\circ$. The uncertainties in the allowed values of $\beta_{pt}$ arise from the uncertainties in the experimental data. Calculating the widths of these decays of the $\eta_2(1645)$ will allow us to narrow down the value of the mixing angle further. The values of the decay widths, given in Table \ref{etatab}, show that the positive mixing angle underestimates the width of the $\eta_2(1645)$ by nearly a factor of $3.5$. Assuming that the $a_2\pi$ channel is the dominant channel for the decay of the $\eta_2(1645)$, the sum of its width along with that of the $K^\ast K$ channel must be close to the total width of the $\eta_2(1645)$. This sum comes out be $\approx 198\pm 15$ MeV for the negative mixing angle and $\approx 58\pm 5$ MeV for the positive angle. These observations hint that the isoscalar mixing angle in the $2^{-+}$ sector must be negative and close to $45^\circ$, consistent with the earlier study reported in Ref. \cite{Koenigstein:2016tjw}.\label{ptangle}
    \item The ratios of the PWAs for the above discussed decays of the $\eta_2(1645)$ are listed in Table \ref{eta2tabrat}. These ratios are independent of the mixing angle. It can be seen from the table that the ratios of PWAs have the same behavior as of those for the decays of the $\pi_2(1670)$. The $D-$waves are less pronounced in the decay of the $\eta_2(1645)$ to the $a_2\pi$ compared to the case of $\pi_2(1670)\to f_2\pi$ as the $\eta_2(1645)$ is marginally lighter than its isovector sibling and the decay product is slightly heavier than $f_2(1270)$ resulting in the 3-momentum carried by the $a_2\pi$ being smaller. But, in the vector mode of the decay, the $F/P-$ratio is comparable to that of the $\pi_2(1670)\to K^\ast K$, as the 3-momenta are nearly the same.
    \item We plot the ratio of the widths of the $\eta_2(1870)\to a_2(1320)\pi$ and $\eta_2(1870)\to f_2(1270)\eta$ decays in the Fig. (\ref{eta2hrat}). The value of this ratio was reported as $1.7\pm 0.4$ in the PDG \cite{Zyla:2020zbs}. From the Fig. (\ref{eta2hrat}), we see that, for this ratio to be small, the mixing angle must be close to zero. It should be noted that, a conventional $\bar{q}q$ model of the $\eta_2(1870)$ predicts the $a_2\pi$, $f_2\eta$, and the $K^\ast K$ channels to be dominant and the mixing angle to be close to zero \cite{Li:2009rka}.\label{ptangle0}
    \item As shown in the Fig. (\ref{eta2hrat}), the local and nonlocal interactions taken separately contribute nearly identically. However, when combined, the width of the $f_2(1270)\eta$ channel (which appears in the denominator) becomes very small leading to a large ratio except when the mixing angle is very small. When the mixing angle takes the values mentioned in point \ref{ptangle} above, the decay widths of the three channels of $\eta_2(1870)$ become significantly smaller than its total width. Specifically, the width of the $f_2(1270)\eta$ channel becomes very close to zero (see Table \ref{etatab}). On the other hand, if the mixing angle is taken to be small and non-zero ($\beta_{pt}=-1.17^\circ$ or $\beta_{pt}=1.42^\circ$), then the width of the $\eta_2(1645)\to a_2\pi$ decay is approximately $368$MeV, which is nearly twice the total width of the $\eta_2(1645)$. Thus, it appears from our analysis that the heavier sibling of the $\eta_2(1645)$ cannot have a mass close to the mass of the $\eta_2(1870)$. This indicates that the $\eta_2(1870)$ is not a member of the $2^{-+}$ $\bar{q}q$-nonet, consistent with the earlier analyses \cite{Anisovich:2010nh,Page:1998gz,Barnes:1996ff,Isgur:1984bm}.
    \item Finally, we compare our results with the results from the $^3P_0$ model (Table \ref{com3P0}). We observe that according to the $^3P_0$ model, all the allowed partial waves ({\it i.e,} the $S-$, $D-$, and the $G-$waves) interfere constructively in the tensor mode of the decay of the $2^{-+}$ mesons. However, this is in contrast with the experimental observations, where, the $D-$waves interfere destructively with the $S-$waves as indicated by the negative $D/S-$ratio. Unfortunately, no information is available about the nature of the $G-$waves. But for this difference in the sign of the ratios, we find that our value for the $\pi_2(1670)\to f_2(1270)\pi$ decay agrees very well with that from the $^3P_0$ model. However, for the other two decays we observe large deviations (factors of $1.5$ and $4$) from the values of the $^3P_0$ model. In the vector decay mode, our values agree fairly well with those of the $^3P_0$ model, assuming similar errors in both the sets of values.
\end{enumerate}

\subsection{Effects of form factors}
The virtual cloud of quarks and antiquarks surrounding the constituent quarks and/or anti-quarks enhance their masses and contribute significantly to the charge radii of the hadrons and hence give rise to finite sizes of hadrons \cite{Povh:1990ad,Lutz:1990zc}.
The non-point-like nature of the mesons brings forth the question of whether a tree-level analysis captures the physics of their decays effectively. In the absence of a fundamental theory or a systematic effective theory, we are forced to use empirical form factors to include the finite-size effects on the decays. Various types of form factors have been used in the past to model the structure of the mesons including, but not limited to Gaussian, exponential, multipole, etc. Specifically, the Gaussian form factors have been used in non-relativistic quark models to study the decays and interactions of mesons \cite{Barnes:1996ff,Amsler:1995td}, the interactions between the nucleons \cite{Shastry:2018oix,Vijande:2003gk} as well as in field theoretic models to study line shapes of various mesons \cite{Wolkanowski:2015jtc}. We make use of the Gaussian form factor of the form \cite{Barnes:1996ff,Amsler:1995td}
\begin{align}
    F(k,\beta) &= e^{-\frac{k^2}{12\beta^2}}.
\end{align}
The form factor contributes to the decay width in the form
\begin{align}
    \Gamma_F &= \Gamma~ |F(k,\beta)|^2,
\end{align}
where $\Gamma_F$ and $\Gamma$ are the decay widths with and without form factors respectively. This is based on the assumption that the decay amplitudes ($i\mathcal{M}$) must be modified to $i\mathcal{M}F(k,\beta)$. Thus, the ratios of PWAs are unaffected by the inclusion the form factor. The typical value of $\beta$ used in the quark model calculations ({\it e.g.,} the $^3P_0$ model) is $0.4-0.5$ GeV \cite{Amsler:1995td}. In the present work, we use the value $\beta=0.4$ GeV. We then extract the values of the parameters using the procedure described in the previous subsections. The values of these parameters are listed in Table \ref{parff}.
Since the value of the form factor for non-zero 3-momentum is always less than one, the parameters become slightly larger when form factor is included. However, the new values and the old values overlap significantly. Thus, even though the form factor modifies the decay widths, the change is not drastically large.\par
\begin{table*}[ht]
    \centering
    {\renewcommand{\arraystretch}{1.5}
    \begin{tabular}{cccccc}
    \hline\hline
        $g_c^A$ (GeV) & $g_d^A$ (GeV$^{-1}$) & $\theta_a$ & $g_c^B$ (GeV) & $g_d^B$ (GeV$^{-1}$) & $\theta_{pv}$\\\hline
        $4.08 \pm 0.79$ & $-0.34\pm 0.39$ & $(26.1 \pm 3.0)^\circ$ & $6.67\pm 0.75$ & $-4.58 \pm 0.38$ & $(26.5\pm 3.0)^\circ$\\\hline\hline
        $g_c^{PT}$ (GeV) & $g_d^{PT}$ (GeV$^{-1}$) & $g_v^{PT}$ & $g_t^{PT}$ (GeV$^{-2}$) & &\\\hline
        $41\pm 13$ & $-8.51\pm 3.10$ & $-11.1\pm 1.46$ & $2.83\pm 0.59$ & &\\\hline
    \end{tabular}}
    \caption{Values of the parameters when form factor is included.}
    \label{parff}
\end{table*}
The need to include nonlocal/tensor interactions to explain the properties of the mesons discussed in this paper tells us that the internal dynamics of the mesons play a crucial role in their decays. Naively speaking, the need for the higher dimension operators indicate the possibility of a scale associated with these processes. Along these lines we would like to note that, the magnitude of the ratios of the parameters $g_v^{PT}/g_t^{PT}$ is approximately $3.92$GeV$^2$ ($\sim 1.5 M_{\pi_2}^2$) for the pseudotensor coupling constants. For the tensor modes, the ratio $|g_c^{PT}/g_d^{PT}|$ is $4.78$GeV$^2$ ($\sim 2M_{\pi_2}^2$) for pseudotensor coupling constants. Similarly, the corresponding ratio in the pseudovector sector is $|g_c^B/g_d^B|$ is approximately $1.52$ GeV$^{-2}$ ($\sim M_{b_1}^2$). From these, we deduce that,  for the pseudotensors $g_v^{PT}\sim M_{PT}^2~\!g_t^{PT}$, $g_c^{PT} \sim M_{PT}^2~\! g_d^{PT}$, and $g_c^B\sim M_B^2~\!g_d^B$ for the pseudovectors, implying that nonlocal interactions play an important a role. 

\section{Summary and Outlook}\label{SnCPWA}
In this work, we have studied the vector decays of the axial-vector, pseudovector, and pseudotensor mesons, and the tensor decays of the pseudotensor mesons. We have derived the partial wave amplitudes for these decays using the covariant helicity formalism. We have demonstrated that the nonlocal interactions play a crucial role in these decays, except in the decays of the axial-vector mesons, where, contact interactions can reproduce the decay widths and the ratio of the PWAs up to a reasonable accuracy. The partial decay widths can be reproduced within the limits of experimental errors and theoretical uncertainties using our approach. 

We also have estimated the mixing angle between the iso-singlets in the $J=1$ sector. The angle of mixing between the axial-vector iso-singlets agrees with the experimentally derived value. But, our model disagrees with the experiments in the pseudovector sector.\par
Similarly, in the decays of the pseudotensor isovectors, the derivative/tensor interactions play a major role and are essential to describe the ratios of the PWAs. We have also studied the isoscalar mixing and we find that the mixing angle must be large and negative ($\approx\!-44^\circ$). Further, we find that the interpretation of the $\eta_2(1870)$ as the heavier partner of the $\eta_2(1645)$ needs further studies. More information about the $\eta_2$ states, in the form of the values of the branching ratios, can help us pin down the mixing angle as well as the nature of the $\eta_2(1870)$.\par
As reported in \cite{Baker:2003jh}, the decay of the $\pi_1(1600)$ into $b_1(1235)\pi$ is know to receive a significant contribution from the $D-$waves. A study along these lines can help in revealing the nature of the hybrid. Also of interest are the $J=1,2$ kaons, which are known to exhibit inter-nonet mixing. Investigation of the partial waves of the kaonic decay can possibly settle the debate on the angle of mixing between these states \cite{Tayduganov:2011ui}.\par

Moreover, in the future one can extend the present study in various
directions, e.g. to higher spin as $J=3$ \cite{jafarzade,wangwang1}, where the data results can be compared to the lattice results \cite{Johnson:2020ilc}, and to baryonic decays. Quite interestingly, the study of PWA is not confined to the strong interaction only. 
Another important future works include the link of PWA to loop effects, thus
going beyond tree-level studied in this work. This can be achieved by taking
into account the widths of the unstable states, both in the initial and the
final states.

In conclusion, partial wave analysis of the decay processes can provide deeper insights into the structure and properties of conventional mesons as well as exotic states and can be of great use in future studies of resonances. 

\section{Acknowledgement}
We are thankful to U. Raha and A. Koenigstein for useful discussions. F. G. and V. S. acknowledge financial support through the Polish National Science Centre (NCN) via the OPUS project 2019/33/B/ST2/00613. F.G. acknowledges also support from the NCN OPUS project no. 2018/29/B/ST2/02576. E.T. acknowledges financial support through the project AKCELERATOR ROZWOJU Uniwersytetu Jana Kochanowskiego w Kielcach (Development Accelerator of the Jan Kochanowski University of Kielce), co-financed by the European Union under the European Social Fund, with no. POWR.03.05.00-00-Z212/18.

\appendix
\section{Unstable states in decay products}\label{specintappe}
Some of the decays discussed in this paper involve unstable final states, {\it e.g.,} $a_1(1260)\to\rho\pi$, $h_1(1170)\to\rho\pi$, and $\pi_2(1670)\to\rho\pi$. Modeling these decays using only the tree level diagrams, {\it prima facie}, does not capture the complete dynamics of the decay process. The instability of the final state can be taken into account by integrating the decay width weighted by the spectral function of the unstable state. Accordingly, for the decay process $A\to BC$ with $B$ unstable, the actual decay width can be written as
\begin{align}
    \Gamma_T &= \int_{s_{th}}^{\infty} ds~ \Gamma_{tree}(s) d_B(s), \label{specint}
\end{align}
where $d_B(s)$ is the spectral function of $B$, $\sqrt{s}=M_B$, and $s_{th}$ is the threshold for the decay of $B$. In spite of this apparent shortcoming, we find that the decay widths do not vary significantly from their tree-level values, as we show using the three decays mentioned earlier.\par
\begin{table}[h]
    \centering
    \begin{tabular}{|c|c|c|}
        \hline
        Decay & Tree-level & Integrated\\
        &  width (MeV) &  width (MeV)\\\hline
        $a_1(1260)\to\rho\pi$ & $420\pm 35$ & $354\pm 30$\\
        $h_1(1170)\to\rho\pi$ & $146\pm 14$ & $142\pm 14$\\
        $\pi_2(1670)\to\rho\pi$ & $80.6\pm 10.8$ & $93.5\pm 13$\\\hline
    \end{tabular}
    \caption{Comparison of the tree-level widths of the three decays with the integrated widths.}
    \label{specinttab}
\end{table}
Typically, the Breit-Wigner form of the spectral function is used to model the unstable state. In this work, we use a different parameterization of the spectral function, called the Sill distribution, which, is normalized to unity and has a built-in threshold \cite{Giacosa:2021mbz}. In all the example decays, the final unstable states is the $\rho-$meson which decays primarily into two pions. Thus, it is sufficient if we use the single channel Sill distribution. The explicit form of distribution function is \cite{Giacosa:2021mbz}
\begin{align}
    d_{B=\rho}(s) &= \frac{1}{\pi} \frac{\tilde{\Gamma}\sqrt{s-s_{th}}}{(s-m_\rho^2)^2+(\tilde{\Gamma}\sqrt{s-s_{th}})^2},
\end{align}
where $\tilde{\Gamma} = \frac{\Gamma_\rho m_\rho}{\sqrt{m_\rho^2-s_{th}}}$, $m_\rho$ is the mass of the $\rho-$meson and $\Gamma_\rho$ is its total width. The decay width mentioned in the integrand of Eq. (\ref{specint}) is given by the Eq. (\ref{decwid1plus}) and Eq. (\ref{PTvecd}) with the mass $M_\rho=\sqrt{s}$. We list the decay widths obtained in the Table \ref{specinttab}. The errors listed are based on the assumption that the fractional errors before and after spectral integration are the same. The widths of the decays do not vary greatly after spectral integration.\par
\begin{table*}[t]
    \centering
    {\renewcommand{\arraystretch}{1.5}
    \begin{tabular}{|c|c|c|c|c|c|c|}
        \hline
         Decay & \multicolumn{3}{c|}{Present Calculation} & \multicolumn{3}{c|}{$^3P_0$ model \cite{Barnes:1996ff}} \\\cline{2-7}
         & $D/S$ & $G/S$ & $F/P$ & $D/S$ & $G/S$ & $F/P$\\\hline
         $a_1(1260)\to\rho\pi$ & -0.062 &  &  & -0.147 &  & \\
         $f_1'(1420)\to K^*K$ & -0.0076 &  &  & -0.026 &  & \\\hline
         $b_1(1235)\to\omega\pi$ & 0.277 &  &  & 0.284 &  & \\
         $h_1(1170)\to\rho\pi$ & 0.281 &  &   & 0.207 &  & \\
         $h_1'(1415)\to K^*K$ & 0.021 &  &  & 0.039 &  & \\\hline
         $\pi_2(1670)\to f_2(1270)\pi$ & -0.18 & 0.0042 &  & 0.185 & 0.0065 & \\
         $\pi_2(1670)\to f_2'(1520)\pi$ & 0.0093 & -7.49$\times 10^{-6}$&  & 0.00697 & 2$\times 10^{-5}$& \\
         $\eta_2(1645)\to a_2(1320)\pi$ & -0.089 & 0.0011 &  & 0.378 & 0.0026 & \\\hline
         $\pi_2(1670)\to\rho\pi$ &  &  & -0.72 &  &  & -0.653\\
         $\pi_2(1670)\to K^*K$ &  &  & -0.447 &  &  & -0.251\\
         $\eta_2(1645)\to K^*K$ &  &  & -0.32 &  &  & -0.193\\\hline
    \end{tabular}}
    \caption{Ratios of the PWAs obtained from the $^3P_0$ model compared to the present calculations.}
    \label{com3P0}
\end{table*}

\section{Comparison with the \texorpdfstring{$^3P_0$}{3P0} model}\label{3P0rev}
In this appendix, we compare our results with those obtained using the $^3P_0$ model. Before presenting the results, we briefly describe the $^3P_0$ model, and list the PWAs given in \cite{Barnes:1996ff}.\par
The $^3P_0$ model is essentially a flux-tube breaking model. In this model, the mesons are assumed to have quark-antiquark pairs with chromoelectric or chromomagnetic flux lines between them. The strong decays of these mesons occur when these flux tubes break and create a quark-antiquark pair. The two quarks and two antiquarks then rearrange into two quark-antiquark pairs which are interpreted as two mesons \cite{Kokoski:1985is}. The Hamiltonian for such a decay is based on the Hamiltonian of the lattice QCD and is given by:
\begin{align}
    \hat{H}|\vec{r}_1,\vec{r}_2\rangle &= \gamma~\!F(\vec{r},\vec{w})~\Psi^\dagger(\vec{R}) \vec{\alpha}\cdot\overleftrightarrow{\nabla} \Psi(\vec{R})|\vec{r}_1,\vec{r}_2\rangle,
\end{align}
where $\Psi(\vec{R})$ is the wavefunction of the $\bar{q}q$ pair created at position $\vec{R}$ with (quark model) quantum numbers $^3P_0$, $\gamma$ is a parameter that captures the strength of flux-tube breaking and, in the wide flux-tube approximation, the function $F(\vec{r},\vec{w}) = 1$ \cite{Geiger:1994kr}. In the non-relativistic limit, the Hamiltonian for the decay $P\to D_1D_2$ reduces to
\begin{align}
    \langle D_1 D_2|\hat{H}|P\rangle &= \gamma\!\int\! \frac{d^3rd^3y}{(2\pi)^{3/2}} e^{i(\vec{p_1}\cdot\vec{r})/2} \Psi_P(\vec{r})\langle\vec{\sigma}\rangle (i\vec{\nabla}_1\nonumber \\
    &\!\!\!\! i\vec{\nabla}_2+\vec{p}_1)\Psi_1^*\left(\frac{\vec{r}}{2}+\vec{y}\right)\Psi_2^*\left(\frac{\vec{r}}{2}-\vec{y}\right),
\end{align}
where the subscripts $P$, $1$, and $2$ represent the parent and the product states respectively, $\langle\vec{\sigma}\rangle$ is the expectation value of the Pauli spin-vector for the $^3P_0$ $\bar{q}q$ pair, and $\vec{p}_1$ is the 3-momentum of the $|D_1\rangle$ state \cite{Barnes:1996ff}. In the above expression, the wavefunctions of the states involved are taken as Harmonic oscillator wavefunctions, which contain the oscillator size parameter $\beta$. The two parameters ($\gamma$ and $\beta$) are fitted to the decay widths of the mesons. We have used value $\beta=0.4$GeV, as given in Ref. \cite{Barnes:1996ff}. Since $\gamma$ is an overall factor multiplying the decay amplitude, the PWAs depend only on the oscillator size parameter. The amplitudes for the decay of the axial-vector and pseudovector mesons to vector and pseudoscalar mesons are given by \cite{Barnes:1996ff}
\begin{align}
    \mathcal{A}_{\ell}(1^{++}\to 1^{--}0^{-+}) &= \Bigg\{ \begin{matrix*}f_S & \ell=0 \\ -\sqrt{\frac{5}{6}} f_D & \ell=2 \end{matrix*}\\
    \mathcal{A}_{\ell}(1^{+-}\to 1^{--}0^{-+}) &= \Bigg\{ \begin{matrix*}-\frac{1}{\sqrt{2}} f_S & \ell=0 \\ -\sqrt{\frac{5}{3}} f_D & \ell=2 \end{matrix*},
\end{align}
where
\begin{align}
    f_S &= \frac{2^5}{\sqrt{3^5}}\left( 1-\frac{2}{9} \frac{k^2}{\beta^2} \right)\\
    f_D &= \frac{2^6}{3^4\sqrt{5}}\frac{k^2}{\beta^2}.
\end{align}
For the decays of the pseudotensor mesons, the amplitudes are given by
\begin{widetext}
\begin{align}
    \mathcal{A}_\ell (2^{-+}\to 2^{++}0^{-+})&= \Vast\{ \begin{matrix*}[l] -\frac{2^6}{\sqrt{3^7}}\left(1 -\frac{5}{18}\frac{k^2}{\beta^2}+\frac{1}{135}\frac{k^4}{\beta^4} \right) & \ell=0 \\ -\sqrt{\frac{2^9 35}{3^{11}}} \frac{k^2}{\beta^2}\left( 1-\frac{4}{105}\frac{k^2}{\beta^2}\right) & \ell=2 \\ -\sqrt{\frac{2^{13}}{3^{11}7}}\frac{1}{5}\frac{k^4}{\beta^4} & \ell=4\end{matrix*}\\\nonumber\\
    \mathcal{A}_\ell(2^{-+}\to 1^{--}0^{-+}) &= \Bigg\{ \begin{matrix*}[l]\frac{1}{2}f_P & \ell=1\\ -\sqrt{\frac{7}{5}} f_F & \ell=3
    \end{matrix*},
\end{align}
\end{widetext}
where
\begin{align}
    f_P &= \frac{\sqrt{2^{13}}}{3^4}\frac{k}{\beta}\left( 1-\frac{2}{15}\frac{k^2}{\beta^2}\right)\\
    f_F &=- \frac{2^6}{\sqrt{3^9 35}}\frac{k^3}{\beta^3}.
\end{align}
These amplitudes are related to the decay amplitude through the relation,
\begin{align}
    \mathcal{M}&= \frac{\gamma}{\sqrt{\beta \pi^{1/2}}}A_\ell e^{-k^2/16\beta^2},
\end{align}
where the final exponential factor represents a form factor for the decay process. The ratios of the PWAs obtained using these expressions are listed in Table \ref{com3P0}.

\end{document}